\documentclass[preprint2]{aastex62} 

\graphicspath{{./}{figures/}}

\accepted{December 26, 2018, for publication in ApJ}
\usepackage{gensymb}
\usepackage{amsmath}
\usepackage{amssymb}
\usepackage{graphicx}
\usepackage{subfigure}
\usepackage{wasysym}
\usepackage{url}
\usepackage[figure,figure*]{hypcap}
\providecommand{\e}[1]{\ensuremath{\times 10^{#1}}}
\shorttitle{Detailed X-ray Mapping of the Ejecta and CSM in SNR G$292.0+1.8$}
\shortauthors{Bhalerao et al. 2018}
\begin{document}
\title{DETAILED X-RAY MAPPING OF THE SHOCKED EJECTA AND CIRCUMSTELLAR MEDIUM IN THE GALACTIC CORE-COLLAPSE SUPERNOVA REMNANT G292.0+1.8}
\author{Jayant Bhalerao}
\affil{Department of Physics, University of Texas at Arlington, P.O. Box 19059, Arlington, TX 76019, USA\\}
\email{jayant.bhalerao@mavs.uta.edu}

\author{Sangwook Park}
\affil{Department of Physics, University of Texas at Arlington, P.O. Box 19059, Arlington, TX 76019, USA\\}

\author{Andrew Schenck}
\affil{Department of Physics, University of Texas at Arlington, P.O. Box 19059, Arlington, TX 76019, USA\\}

\author{Seth Post}
\affil{Department of Physics, University of Texas at Arlington, P.O. Box 19059, Arlington, TX 76019, USA\\}

\author{John P. Hughes}
\affil{Department of Physics and Astronomy, Rutgers University, 136 Frelinghuysen Road, Piscataway, NJ 08854-8019, USA\\}

\begin{abstract}
G292.0+1.8 (G292) is a young ($\sim$3000 yr), Galactic textbook-type core-collapse supernova remnant (CCSNR). It is characterized by X-ray, optical and infrared emission from ejecta and circumstellar medium (CSM) features, and contains a pulsar (PSR J1124-5916) and pulsar wind nebula that have been observed in X-rays and radio. Previous studies have revealed a complex, dynamically evolving, oxygen-rich remnant, a striking relic from the explosion of a massive star. Here, using our deep (530 ks) \textit{Chandra} ACIS data, we present high spatial-resolution maps (based on a regional grid size of a few arcsec) of the shocked CSM and metal-rich ejecta in G292. We make the first {\it Chandra}-detection of Fe-rich ejecta in G292. We identify the X-ray counterpart of the northern equatorial belt, a component of a ring-like CSM structure identified earlier in the infrared band. We show the detailed spatial distributions of ejecta enriched in O, Ne, Mg, Si, S and Fe. We find that the bulk of the Si, S and Fe-rich X-ray-emitting ejecta are located in the northwestern hemisphere of the remnant, opposite to the pulsar's projected angular displacement to the southeast from the SNR's center. This suggests that the pulsar's kick may have originated from gravitational and hydrodynamic forces during an asymmetric explosion, rather than from anisotropic neutrino emission. Based on abundance ratios and our estimated CSM and ejecta masses, we constrain the progenitor mass to 13 M$_{\astrosun}$ $\lesssim$ M $\lesssim$ 30 M$_{\astrosun}$

\end{abstract}


\section {\label {sec:intro} INTRODUCTION}

Oxygen, a key component of life, is produced by nuclear fusion in the cores of massive stars and released into the interstellar medium when these stars explode as supernovae (SNe). G292.0+1.8 (G292 hereafter) is one of three known Galactic oxygen-rich core-collapse supernova remnants (CCSNR, \citealt{goss1979}). The other two in this group are Cassiopeia A (Cas A) and Puppis A (e.g., \citealt{ghavamian2005} and references therein). These remnants show optical emission dominated by forbidden lines of oxygen from high-speed SN ejecta. The oxygen is produced by nuclear burning processes in the interiors of the massive progenitor stars that gave rise to these remnants. Oxygen and other nucleosynthesis products are expelled into interstellar space as SN ejecta when these stars explode as CCSN. Among the three known Galactic oxygen-rich remnants, only G292 shows all the $\textit{textbook-type}$ features of a CCSNR: metal-rich ejecta, shocked circumstellar medium (CSM), a rotation-powered neutron star (NS or pulsar, PSR J1124-5916) and pulsar wind nebula (PWN) detected both in X-rays and in radio (\citealt{hughes2001}, \citeyear{hughes2003}; \citealt{camilo2002}; \citealt{gaensler2003}). In contrast, Puppis A is dominated by emission from shocked gas with a low-abundant CSM and/or ISM-like composition (\citealt{hwang2008}; \citealt{katsuda2008}; \citealt{luna2016}). The NS in Puppis A has a weak magnetic field and is radio-quiet (\citealt{gotthelf2009}). Cas A differs in having ejecta that are unusually low in Ne and Mg abundance, and are dominated by Si, S and Fe (\citealt{vink1996}; \citealt{dewey2007}). The NS in Cas A does not pulsate and it is not surrounded by a PWN (e.g., \citealt{pavlov2000}; \citealt{vink2008}). G292's textbook features make it very useful for studying the evolutionary processes of stars, massive-enough to eventually explode as ``standard" CCSNe and create normal pulsars.

G292 is the result of an unrecorded SN explosion in an area of the southern sky marked by the bright constellation {\it Centaurus}. Its age of $\sim$3000 yr was estimated from the observed expansion rates of fast-moving optical ejecta knots in the optical band (\citealt{ghavamian2005}; \citealt{winkler2009}) and from the Sedov interpretation of its X-ray shell emission spectrum (\citealt{gonzalez2003}). This age estimate is consistent with the characteristic spin-down age of G292's pulsar (2900 yr, \citealt{camilo2002}). The distance to G292 has been estimated to be $\gtrsim$6 kpc using H I absorption measurements (\citealt{gaensler2003}). G292 has an angular size of $\sim$9$\arcmin$ in X-ray (\citealt{park2007}) and in radio (\citealt{gaensler2003}), which corresponds to a diameter of $\sim$16 pc at $d = 6$ kpc. 

At X-ray wavelengths, G292 exhibits a rich, and intricate pattern of ejecta and CSM structures. The ejecta form networks of knots and filaments distributed over the face of the entire remnant. Superimposed on the ejecta is the shocked CSM, which manifests as several distinctive morphological structures in X-rays: the equatorial belt, thin circumferential filaments, and the outermost diffuse, spectrally soft emission (\citealt{gonzalez2003}; \citealt{park2002}, \citeyear{park2004}, \citeyear{park2007}). The equatorial belt is a dense, belt-like feature running along the SNR's ``equator." It has been observed in the optical band (\citealt{ghavamian2005}), in infrared (\citealt{lee2009}; \citealt{ghavamian2012}; \citealt{ghavamian2016}), and in X-rays (e.g., \citealt{tuohy1982}; \citealt{hughes2001}; \citealt{park2002}, \citeyear{park2004}, \citeyear{park2007}). The thin circumferential filaments form narrow arcs of spectrally soft emission along the outer boundary of the SNR (\citealt{gonzalez2003}; \citealt{park2002}, \citeyear{park2007}). Finally, the shocked CSM forms a diffuse, spectrally soft border along the outer edge of the SNR which marks the region where the forward shock (FS) is interacting with the red supergiant (RSG) winds of the progenitor (\citealt{park2007}; \citealt{lee2010}). The suggested origins for these CSM emission features may include relic structures from the outer atmosphere of a rotating progenitor star, and/or from binary interactions (e.g., \citealt{chevalier1992}; \citealt{chita2008}; \citealt{morris2009}; \citealt{smith2013}).

The ejecta in G292 are characterized by high velocity ($v_{radial}$ ${\gtrsim}$ 1000 km s$^{-1}$) knots and filaments, also known as fast-moving knots or FMKs (\citealt{ghavamian2005}; \citealt{bhalerao2015}). The bulk of the shocked metal-rich ejecta in G292 is dominated by O, Ne and Mg (e.g., \citealt{park2004}). Relatively weak Fe K-shell line emission from the Fe-rich ejecta gas has been detected in \textit{Suzaku} data (\citealt{kamitsukasa2014}; \citealt{yamaguchi2014}), however the poor angular resolution of \textit{Suzaku} ($\sim$2$\arcmin$ HPD, \citealt{mitsuda2007}) hindered its accurate spatial mapping. The location of the reverse shock (RS) close to the outer edge of the PWN in G292 (\citealt{gaensler2003}; \citealt{bhalerao2015}) suggests that the interaction of the RS with the central ejecta may have recently started, therefore the bulk of the innermost ejecta, representing explosive nucleosynthesis products such as Fe, may not have been heated by the RS yet. Unlike Cas A, significant mixing and overturning of ejecta in G292 may not have occurred (\citealt{park2004}; \citealt{ghavamian2012}).

Previous studies have sampled a limited number of ejecta regions using \textit{Chandra} data (\citealt{gonzalez2003}; \citealt{park2004}, \citeyear{park2007}). More extensive regions of G292 have been analyzed using a regional grid (\citealt{yang2014}), however, this study was based on {\it Chandra} Advanced CCD Imaging Spectrometer-S3 (ACIS-S3) data with incomplete coverage of the SNR because of the detector's smaller field of view (8.3$\arcmin$$\times$8.3$\arcmin$, \citealt{weisskopf2005}). In this paper, we use data collected with the \textit{Chandra} ACIS-I array, which has a larger field of view (17$\arcmin$$\times$17$\arcmin$) and covers the entire SNR. Furthermore, the ACIS-I data we use here has an order of magnitude longer exposure than the ACIS-S3 data. 

Evidence for a strong link between asymmetric SN explosions and ``NS-kicks" (forces imparted to the NS during the SN explosion) is emerging (e.g., \citealt{janka2017} and references therein). In G292, evidence for an asymmetric explosion has been suggested. Si-rich ejecta gas is observed mainly in the north and northwest of the SNR (\citealt{park2002}; \citealt{yang2014}). Radial velocity measurements detect a significantly larger number of blueshifted knots compared to redshifted ones (\citealt{ghavamian2005}; \citealt{bhalerao2015}). Also, the radial velocity magnitudes of the blueshifted knots are generally higher than those of the redshifted ones (\citealt{bhalerao2015}). Oxygen-rich ejecta filaments in the optical band show higher proper motions in the north-south direction than in the east-west direction (\citealt{winkler2009}). 

The detailed distribution of Fe-rich ejecta in G292 is unknown. Fe is a key explosive nucleosynthesis product, produced in the deepest layers of the SN, and its spatial distribution, especially asymmetric patterns, would be crucial for revealing the nature of the CCSN explosion (\citealt{woosley2002}; \citealt{thielemann2007}; \citealt{maoz2017}).

Here, using our deep \textit{Chandra} ACIS-I data, we perform a detailed spectroscopic analysis of the entire remnant. This study is a direct expansion of our earlier studies of G292 based on the same \textit{Chandra} data (\citealt{park2007}; \citealt{lee2010}). In this paper, we reveal G292's structure, morphology and distribution of the CSM and ejecta in unprecedented detail. For the first time, we provide a high-spatial-resolution map revealing the distribution of Fe-rich ejecta in G292. We also provide spatial distribution maps for O-, Ne-, Mg-, Si- and S-rich ejecta, and their thermodynamic parameters including the electron temperature and ionization timescale. We discuss these results in the context of recent CCSN hydrodynamic models, and their implications in understanding the nature of CCSN explosions.

\pagebreak
\section{\label{sec:obs} OBSERVATIONS \& DATA REDUCTION}
Our G292 observation was performed between 2006 September 13 and 2006 October 20 using the ACIS-I array. We used the ACIS-I array since it has a large field of view (17$\arcmin$$\times$17$\arcmin$, \citealt{weisskopf2005}) and can cover the entire SNR (angular size $\sim$9$\arcmin$, \citealt{park2007}). The observation consisted of six individual ObsIDs with exposure times ranging from $\sim$40 ks to 160 ks (Table 1). The aim point was close to the position of the pulsar (PSR J1124-5916) at R.A. (J2000.0) = $11^{h} 24^{m} 39^{s}.1$, decl. (J2000) = -59\degree 16\arcmin 20\arcsec~(\citealt{camilo2002}; \citealt{hughes2003}). We processed the data with standard data reduction methods for grade and hot pixel filtering using the Chandra Interactive Analysis of Observations (CIAO, \citealt{fruscione2006}), version 4.7, with calibration database CALDB 4.6.7 . No significant background flaring was observed. The processed data had a total effective exposure time of $\sim$509 ks. We note that after we had completed our analysis for this work using our 2006 observation, additional ACIS-I data of G292 (with a total exposure of $\sim$300 ks, taken in 2016) became available in the archive. Considering the time-dependent quantum efficiency degradation of the ACIS-I detector, we estimate that although these additional data would increase the total number of photons by $\sim$40\%, the added signal-to-noise will not alter our scientific conclusions. Thus, we do not include these additional \textit{Chandra} archival data of G292 in this work.

\section{\label{sec:result} ANALYSIS \& RESULTS}
\subsection{Characterization of the Outermost Shocked CSM}
We divided the remnant into a sub-regional grid using our adaptive mesh method (\citealt{schenck2016}). This technique adaptively divides the SNR into small rectangular sub-regions, to contain a certain minimum number of counts per sub-region. We used the 0.3-8 keV band image of G292 to apply this method. To perform a statistically significant spectral analysis of each regional spectrum, we set each individual region to contain at least 5000 counts in the 0.3-8 keV band. We find that several thousand counts are generally required to obtain measurements of spectral model parameters, including elemental abundances, to within $\sim$50\% uncertainties. Constrained measurements with uncertainties $\lesssim$ 50\% allow reliable characterization of the physical and chemical properties of the X-ray emitting gas in SNRs. From our initial spectral analysis of a small number of regions in G292, we found that $\sim$5000 counts were required for such measurements. Our adaptive mesh results in 2147 regions (Fig. 1a). The sizes of the regions range from $\sim$ 2\arcsec$\times$4\arcsec~for regions enclosing the bright filaments, to $\sim$30\arcsec$\times$50\arcsec~for regions representing the faint outer boundary of the SNR. The average angular area of the regions is $\sim$115 arcsec$^2$.

To estimate mean elemental abundances in the shocked CSM in G292, we extracted X-ray spectra from several representative CSM emission regions identified in previous works (e.g., \citealt{lee2010}; the spectrally-soft red diffuse regions near the outermost boundary of the SNR in Fig. 1b). We fitted these regional spectra using the absorbed non-equilibrium ionization (NEI) plane-parallel shock model (\citealt{borkowski2001}) with varied abundances (\textit{vpshock}, NEI version 3.0 with updated ATOMDB (\citealt{smith2001}; \citealt{badenes2006}; \citealt{foster2012}) in XSPEC (\citealt{arnaud1996}) and ISIS, (\citealt{houck2000}). We varied the foreground column ($N_{H}$), normalization, electron temperature ($kT$, where $k$ is the Boltzmann constant and $T$ is the temperature), ionization timescale ($n_{e}t$, where $n_{e}$ is the postshock electron number density and $t$ is the time elapsed since the passage of the shock), and elemental abundances for O, Ne, Mg, Si, S and Fe. We fixed the abundances of all other elements to solar values (\citealt{anders1989}). The measured abundances and thermodynamic parameters did not show significant variation among the regions (within statistical uncertainties), therefore we took the average values to represent the CSM emission. These average values are consistent with previous estimates (\citealt{lee2010}) and are listed in Table 2. 

\subsection{Origin of the Regional Emission (Ejecta vs. CSM) }
To characterize the emission across the remnant, we extracted source emission spectra from all 2147 regions. We subtracted the background emission using spectra extracted from source-free regions outside the remnant. We used a total of four circular background regions located beyond the outer boundary of the SNR, with one background region located on each of the four CCDs of the ACIS-I detector. We combined these four regions to characterize the ``average" background spectrum. We fit spectra for all the individual regions with an NEI plane shock model. For regions projected within or near the PWN, the contribution in the observed regional spectrum from the nonthermal synchrotron radiation from the PWN may be significant. Therefore, we added a power-law component in our spectral model fits for $\sim$310 regional spectra generally projected against the spectrally-hard central PWN feature (\citealt{hughes2001}; \citealt{park2007}). The locations of these regions are outlined with a blue dashed-curve in Fig. 1c. For our spectral model fits, the parameters we varied are: $N_{H}$, normalization, $kT$, $n_{e}t$ and the abundances for O, Ne, Mg, Si, S and Fe. We fixed the abundances of all other elements to solar values (\citealt{anders1989}).

We classified regions as \textit{ejecta-dominated} if the measured abundance for any fitted elemental abundance exceeded our measured CSM abundance values (listed in Table 2) by more than a 3$\sigma$ confidence level. Using this method, we identified $\sim$1400 \textit{ejecta-dominated} regions (white regions in Fig.~1c), and $\sim$700 \textit{CSM-dominated} regions (brown regions in Fig.~1c). In the central regions of the SNR, $\sim$150 \textit{CSM-dominated} regions and $\sim$160 \textit{ejecta-dominated} regions are affected by synchrotron continuum emission from the PWN (these regions are outlined by a blue dashed-curve in Fig.~1c). These regions show a relatively high continuum flux in the 2-7 keV band (${\sim}$ 10\%--60\% of the total flux). They yielded fits with $\chi^2/\nu > 2$ and/or unrealistically high electron temperature ($kT$ $\sim$ 3--10 keV) when no power-law component was included in the spectral model. After we added the power-law component, these PWN regions yielded improved fits with $\chi^2/\nu$ $\sim$ 0.9 to 1.5, and $kT$ $\lesssim$ 0.7 keV. The best-fit photon index is $\Gamma$ $\sim$ 1--3 for these PWN regions, which is generally consistent with values estimated for G292's PWN regions in the literature (\citealt{gonzalez2003}). Overall, our approach yields statistically acceptable fits ($\chi^2/\nu < 2$) for over 99\% of the 2147 regions comprising the SNR (Fig. 5a). In Fig. 2, we show an example spectrum of an \textit{ejecta-dominated} region (region ``E" in Figs. 1a \& 1c), and a \textit{CSM-dominated region} (region ``C" in Figs. 1a \& 1c). Both spectra in Fig. 2 were fitted with an NEI plane shock model assuming our measured average CSM abundances (Table 2). The \textit{ejecta-dominated} region gives a poor fit to this model ($\chi^2/\nu$ = 6.6) due to excessive line fluxes originating from overabundant ejecta gas (primarily from the Ne K and Mg K lines at $E$ $\sim$ 1.0 keV and $\sim$1.35 keV respectively). The CSM-dominated region, on the other hand, gives a good fit ($\chi^2/\nu$ = 1.2).

\subsection{Spatial Distribution of the Elemental Abundances and NEI Plasma Parameters}
We show the measured abundance distributions of O, Ne, Mg, Si, S and Fe for the entire remnant in Fig. 3. For comparisons, we also show atomic line \textit{equivalent width image} (EWI) maps for these elements in Fig. 3. We position these maps next to each other to allow for easier side-by-side, element by element, comparisons of regional abundances and emission line strengths. We created the EWI maps using published methods (e.g., \citealt{hwang2000}; \citealt{park2002}; \citealt{schenck2014}). The line and continuum band energies used for making these maps are listed in Table 3. We binned the images by $2\times2$ pixels, and adaptively smoothed them before performing the EWI calculations. These EWI maps are updated versions of those we had published earlier (\citealt{park2002}). Our previous EWI maps were based on $\sim$43 ks of $\textit{Chandra}$ data taken on the ACIS-S3 wherein parts of the remnant were not imaged due to the small field of view of the ACIS-S3. Our new EWI maps utilize over an order of magnitude higher count statistics than our previous maps. These higher photon counts allow us to rebin the subband images with a significantly smaller number of pixels in our new EWI maps, as compared to our previous maps. This binning advantage helps in effectively creating a higher spatial resolution in our new EWI maps. Our new EWI maps also cover the entire SNR. In this work, we provide EWI maps that have not been published before, namely those for Mg XII, S XV and Fe K (Fe He$\alpha$ or Fe XXV). In our EWI maps, regions with strongly enhanced EWs are generally coincident with highly overabundant regions in our elemental abundance maps (Fig. 3), indicating that the strong line fluxes are primarily caused by the presence of overabundant ejecta gas. In Fig. 3, we label the EWI maps based on the most prominent X-ray emission lines observed for SNR plasmas (e.g., \citealt{vink2017}).

In Fig. 4, we show abundance-ratio maps (for the \textit{ejecta-dominated} regions) of O/(Si+S+Fe), Ne/(Si+S+Fe), Mg/(Si+S+Fe), Si/(O+Ne+Mg), S/(O+Ne+Mg) and Fe/(O+Ne+Mg) that highlight the enhancements of lighter O-group elements and the heavier Si, S and Fe relative to each other. Figs. 3i--n and Figs. 4d--f illustrate the anti-alignment between the pulsar and the heavier Si-, S- and Fe-rich ejecta, with the heavier ejecta clearly localized in the north-northwestern regions of the SNR opposite to the projected location of the pulsar (see Section 4.4).  

In Fig. 5, we show the spatial distributions of the NEI spectral parameters: $N_{H}$, $kT$, $n_{e}t$, the {\it pseudo} electron density ($n_{EM}$), and the {\it pseudo} thermal pressure ($P_{EM}$). We calculated the pseudo density as $n_{EM} = \sqrt{EM/V}$, where $EM$ is the volume emission measure and $V$ is the emission volume (described in Section 4.2). We calculated the pseudo thermal pressure as $P_{EM} = n_{EM} kT$. We label these quantities as ``pseudo" since they are derived from the projected density and temperature of the SNR gas, and not the actual 3D distributions of these quantities within the remnant.

To reveal the detailed radial structure in the ionization state and ejecta abundances, we constructed radial profiles for $kT$, $n_{e}t$, and the abundances of O, Ne, Mg, Si, S and Fe along the NW and SE directions. To effectively study such radial structures, we selected the NW sector of the SNR, where the layering structure of abundances and the progressively ionized nature of the ejecta along the radial distance from the SNR center have been suggested (\citealt{park2002}). For comparisons, we also performed a similar study in the opposite area of the SNR (the SE region). This ``axis'' along NW-SE is also intriguing in light of its alignment with the projected pulsar kick direction (Figs. 3, 4, Section 4.4). For these radial profiles, we divided the ejecta regions into radial intervals of $\sim$15\arcsec. We show these regions in Fig. 6a, and the radial profiles in Figs. 6b--i.

\subsection{Fe-Rich Emission Regions}
We note that the Fe-abundances (Fig. 3k) were measured primarily based on the Fe L-line complex ($E$ $\sim$ 0.7-1.2 keV), since the Fe K-shell line (at $E$ $\sim$ 6.6 keV) is weak or undetectable in most individual regional spectra. We fitted the spectra extracted from Fe-rich regions, with a plane-parallel shock model, with the Fe-abundance fixed at our measured average CSM value (0.13$\times$ solar, Table 2), while varying the abundances of O, Ne, Mg, Si, S, Ar and Ca. The abundance of Ni was tied to Fe. We varied the $N_{H}$, normalization, $kT$ and $n_{e}t$. These model fits resulted in strong residuals at \textit{E} $\sim$ 1.2 keV, consistent with excess emission from the Fe L complex (e.g., \citealt{vink2012};  \citealt{kamitsukasa2014} and references therein). We show an example spectrum with model fits for such a region (Region A in Fig. 7) in Figs. 8a \& b. For comparisons, we show an example spectrum of an ejecta region with low Fe abundance, which shows no significant residuals relative to the best-fit model at \textit{E} $\sim$ 1.2 keV (Region B in Fig. 7) in  Figs. 8c \& d.

The Fe-enhancements are supported by our detection of Fe K-shell line in the integrated spectrum of larger areas over the Fe-enhanced regions with significantly higher photon statistics ($\sim$10$^{6}$ counts in the 0.3-8.0 keV band). We show these large regions in Fig. 7 (regions 1 and 2) and their spectrum in Fig. 9a. For comparisons, the spectrum of an ``Fe-poor"  region, (region 3 in Fig. 7, containing $\sim$1.7\e{6} counts in the 0.3-8 keV band) shows significantly weaker Fe K-shell line emission (Fig. 9b). We estimate a 6$\sigma$ confidence level detection of the Fe K-shell line in the Fe-rich region, while the Fe K-shell line detection in the Fe-poor region's spectrum is marginal (at a 3$\sigma$ confidence level). 

We measured the line center energy for the observed Fe K-shell line in G292 by Gaussian-fitting of the spectrum extracted from combining $\sim$30 Fe-overabundant northern regions. We varied the line center energy, line width and normalization in the Gaussian model. We measure a line center energy of 6.62~$\pm$0.08 keV (90\% confidence range) for the Fe K-shell emission line (Fig. 10). Our measured Fe K-shell line center energy agrees with earlier \textit{Suzaku} measurements (\citealt{kamitsukasa2014}; \citealt{yamaguchi2014}), and is consistent with the detected values for other CCSNRs (\citealt{yamaguchi2014}).

\section{\label{sec:disc} DISCUSSION}

\subsection{Spatial Structure of Ejecta Elements and NEI Plasma Parameters}
The ejecta in G292 show striking differences between the spatial distributions of O, Ne and Mg (representing hydrostatic nucleosynthesis products), and Si, S and Fe (tracers for explosive nucleosynthesis, especially Fe). While O-, Ne-, and Mg-rich ejecta are widely scattered across the remnant, they appear to be particularly enhanced in the NW and SE quadrants (Figs. 3a--c). This preferred distribution is brought out in our O/(Si+S+Fe), Ne/(Si+S+Fe) and Mg/(Si+S+Fe) abundance ratio maps (Figs 4a--c). In contrast to the abundances of the O-group elements, the abundances of the heavier elements, Si, S and Fe, are enhanced almost exclusively in the NW regions (Figs. 3i--n). This enhancement is also revealed in our Si/(O+Ne+Mg), S/(O+Ne+Mg) and Fe/(O+Ne+Mg) abundance ratio maps (Figs. 4d--f).

X-ray line emission for O, Ne, Mg, Si, S and Fe, shown in our line EW maps, reveals distributions that are generally consistent with our abundance maps (Fig. 3). A similarity between the spatial distribution patterns for the line EWs for O, Ne IX and Mg XI is observed (Figs. 3d--f). These line EWs are generally enhanced close to the projected position of the RS.\footnote{The projected position of the RS at $r_{RS}$ $\sim$ 130\arcsec~ and CD at $r_{CD}$ $\sim$ 220\arcsec~are 1D approximations, estimated based on G292's kinematic structure in X-rays (\citealt{bhalerao2015}). $r$ is the angular radius measured from the optical expansion center of the remnant at RA (2000.) = 11$^h$, 24$^m$, 34$^s$.4, and decl. (2000.) = -59$^{\circ}$, 15\arcmin, 51\arcsec (\citealt{winkler2009}). The actual 3D positions of the RS and CD could differ in different parts of the remnant.} 

In the SE, the line EWs for O, Ne IX and Mg XI are enhanced in an area that is coincident with the optical ``spur" (\citealt{ghavamian2005}). The spur is a crescent-shaped structure dominated by low velocity (v $\sim$ 100 km s$^{-1}$, \citealt{ghavamian2005}), dense, clumpy ejecta that are radiatively cooling to emit in the optical band (e.g., [O III], \citealt{ghavamian2005}; \citealt{winkler2006}) and in infrared (e.g., [O IV], \citealt{ghavamian2012}). While the optical and X-ray emitting ejecta in the SE may not precisely share  the same spatial locations throughout the remnant, probably some of the low-density, high velocity X-ray ejecta have mixed with the higher density optical ejecta to produce the observed X-ray emission of O, Ne IX, and Mg XI from the southeastern regions of the SNR. Similar spatial correlations between optical and X-ray emission in G292 have been observed before (\citealt{park2002}, \citeyear{park2007}; \citealt{ghavamian2005}; \citealt{bhalerao2015}).

Line emission for Ne X and Mg XII, on the other hand, is not evident in the SE, and it peaks at larger radial distances in other parts of the remnant, closer to the contact discontinuity (CD) and even extending to the outer boundary in some areas (Figs. 3g \& 3h). This radially ``layered" emission pattern is particularly emphasized in the NW. These high EW-regions in the NW are spatially coincident (in projection) with overabundant ejecta gas (Fig.~3). A similarly layered EW distribution for Ne IX and Ne X lines in the NW regions of G292 have been reported in previous work (\citealt{park2002}). Based on our extensive spectral analysis, we show radially increasing ionization timescale values for the shocked ejecta gas (from the RS to the CD) in the NW (Fig. 6c). This radial distribution of the ionization timescale supports the suggested progressive ionization of the ejecta gas by the RS on its migration towards the center of the remnant. A similar progressive ionization structure for O- and Ne-rich ejecta gas was observed in the young (age $\sim$ 1000 yr) O-rich SNR 1E 0102.2--7219 (\citealt{gaetz2000}; \citealt{flanagan2004}; \citealt{alan2018}). The weak EWs for the highly-ionized Ne X and Mg XII in the SE (Figs. 3g \& 3h) are consistent with the generally lower values of the ionization timescale there ($n_{e}t$ $\sim$ 5\e{11} cm$^{-3}$ s) compared to those in the NW ($n_{e}t$ $\sim$ 10$^{12}$ cm$^{-3}$ s). Regions in the NW located just within our identified CD show the highest $n_{e}t$. These high $n_{e}t$ regions are followed by regions where $n_{e}t$ decreases radially inwards towards the RS (Figs. 5d \& 6c). 

In our radial profiles, O, Ne and Mg peak closer to the CD at about an angular distance of 170$\arcsec$-200$\arcsec$ from the center of the SNR. On the other hand, Si, S and Fe abundances peak at a smaller radial distance of about 130$\arcsec$-150$\arcsec$ from the SNR's center (Figs. 6d--6i). This provides further evidence for a layered ejecta-structure in the NW, where the classical ``onion-shell" nucleosynthesis-configuration at the core of the massive progenitor appears to have been preserved in the SNR. However, this layering of the ejecta elements is not clearly evident in other directions. For example, abundances for O and Fe peak at about the same radial distance ($r$ $\sim$ 130$\arcsec$) in the SE (blue curves in Figs. 6d--i). This may suggest some mixing of the ejecta in the SE and/or projection effects caused by O-rich regions that are physically located at larger radial distances than the Fe-rich regions, but are projected closer to the SNR center along the line of sight. 

Our electron temperature map shows a general range between 0.5 to 1 keV across the remnant (Fig. 5c), but there are regions in the N and NW with significantly higher electron temperature (\textit{kT} $\sim$ 2-4 keV). These high-temperature regions are also enriched in Fe (Fig. 3k). The hot gas temperature in this region may have helped us effectively detect line emission from Si, S and Fe (especially the Fe K-shell line). The higher electron temperatures in the N and NW are generally consistent with the results reported by \citet{park2007}. 

The absorbing column ($N_H$) does not show significant variations across the remnant (Fig. 5b). The average column density over the entire remnant is 6\e{21} cm$^{-2}$. There are a few regions in the S and SW (near the outermost boundary of the SNR) with higher $N_H$ ($\sim$8\e{21} cm$^{-2}$). Although the higher $N_H$ in the S and SW is not statistically significant, an excess $N_H$ in this part of the SNR cannot be ruled out. This is suggested by the enhanced dust emission observed from these regions in the far-infrared (e.g., \citealt{park2007}; \citealt{ghavamian2016}). There may be superposed H II regions in the SW (as suggested by \citealt{rodgers1960}), or a molecular cloud complex projected along and outside the southwestern boundary of G292 (as suggested by \citealt{ghavamian2016}). Detailed maps and distances to these molecular cloud structures are not known, and the presence of a higher column in the S and SW will need further investigation.

\subsection{Ejecta Mass}
We estimated the ejecta masses for the individual elements based on our estimated volume emission measure (e.g., \citealt{lopez2009} and references therein): 
\begin{equation}
EM_x = n_e n_x V f
\end{equation}
where $\boldsymbol{n_x}$ is the ion number density, $V$ is the volume of the region, $f$ is the X-ray emission filling factor which characterizes the clumpiness of the emission, and the subscript $x$ designates the element or ion being evaluated. To estimate $V$ we assumed a distance of $d$ = 6 kpc to the SNR. We assumed that the ejecta knots and filaments represent localized volumes of X-ray-emitting gas, for which the line of sight depth is of the order of the angular extent of each rectangular region.\\
We calculated the electron number density:
\begin{equation}
n_e = \sqrt{\frac{EM_x B_x X_x N_x}{V}}
\end{equation}
where $B_x$ is the best-fit abundance, $X_x$ is the solar number abundance and $N_x$ is the average number of electrons lost by a given ion. We estimated $N_x$ using a simple approximation for O-rich SNR gas (\citealt{lazendic2006}):
\begin{equation}
N_{x} = 
	\begin{cases}
	Z,\;\;\;\;\;\;\;\;\;\; \text{for}\; Z \leq 9\\
    Z-2,\;\;\;\; \text{for}\; 10 \leq Z \leq 16\\
    Z-10, \;\;     \text{for}\; Z \geq 17
	\end{cases}
\end{equation}
where $Z$ is the atomic number of the element. Using this approximation we get the following $N_x$: O = 8, Ne = 8, Mg = 10, Si = 12, S = 14, Fe = 16. To calculate $N_x$ for Fe, we assumed that the Fe XVII ion is the dominant contributor to the Fe L complex detected at $E$ $\sim$ 0.7-1.2 keV in G292 (e.g., \citealt{kamitsukasa2014}, and this work). We note that assuming the higher ionized state of Fe XXV ($N_x$ = 24), corresponding to the observed Fe K-shell line emission at $E$ = 6.6 keV (e.g., \citealt{vink2012} and references therein, \citealt{vink2017}), does not significantly alter our estimated Fe ejecta masses. The difference in estimated Fe ejecta masses assuming these different ionization states is less than 20\%, and within statistical uncertainties.\\
We calculated the mass of the element in each ejecta region:
\begin{equation}
M_{x} = n_x A_x m_u V f
\end{equation}
where, $n_{x} = n_{e}/N_{x}$, $A_x$ is the atomic mass for the dominant isotope and $m_u$ is the atomic mass unit in grams (1 amu = $1.66\times10^{-24}$ g).

We calculated the total ejecta mass for each element by summing the calculated masses for all ejecta regions. We list our calculated ejecta masses for O, Ne, Mg, Si, S, Fe for the entire remnant, and the total ejecta mass in Table 4. We estimate a total shocked ejecta mass of $\sim$1.29$^{+0.49}_{-0.28}$ $f^{1/2}$ $d^{3/2}_{6}$ M$_{\astrosun}$. In Fig. 11 we show the ``quadrants'' and ``hemispheres'' we use to investigate asymmetries in the ejecta elemental mass distributions. We list the calculated ejecta masses for these quadrants and hemispheres in Table 4.

We estimate that a total of $\sim$0.03$\pm{0.01}$ $f^{1/2}$ $d^{3/2}_{6}$ M$_{\astrosun}$ of Fe ejecta has been shocked by the RS to emit X-rays. This is significantly less than the Fe-yield of $\sim$0.07-0.1 M$_{\astrosun}$ predicted for a wide range of progenitors (M = 13--40 M$_{\astrosun}$), metallicity ($Z$ = 0--0.02) and explosion energy ($E$ = 1--30 \e{51} ergs)(\citealt{nomoto2006}). Similarly, we estimate that $\sim$0.06$\pm{0.01}$ $f^{1/2}$ $d^{3/2}_{6}$ M$_{\astrosun}$ of Si ejecta, and $\sim$0.02$\pm{0.01}$ $f^{1/2}$ $d^{3/2}_{6}$ M$_{\astrosun}$ of S ejecta have been shocked (Table 4). Thus, $\sim$10-80\% of Si, and $\sim$30-80\% of S ejecta may still remain unshocked in the SNR (comparing them with yields predicted for progenitors with mass 13--40 M$_\odot$, $Z$ = 0.02 and $E$ = 1\e{51} ergs, \citealt{nomoto2006}, also see Section 4.5). This suggests that only the outskirts of the central ejecta gas may have been shocked by the RS. This scenario is supported by the undistorted morphology of the PWN, an indication that the RS may have just reached the central regions of the SNR where the Si-, S- and Fe-rich ejecta are expanding (\citealt{gaensler2003}; \citealt{gaensler2006}; \citealt{bhalerao2015}). 

\subsection{Properties of the CSM}
We identified the \textit{CSM-dominated} regions in G292 based on their low abundances estimated by our regional spectral model fits. Our \textit{CSM-dominated} regions in Fig. 1c trace the various CSM features identified in earlier works (see Section 1). These include the thin circumferential filaments and the spectrally soft diffuse emission at the outer boundary. The origin of the thin circumferential filaments is unclear. They may be the result of the blast wave interacting with stellar wind structures produced during late evolutionary stages of the G292 progenitor (e.g. RSG and blue supergiant phases, \citealt{park2002}). The ``inner boundary" of the spectrally soft diffuse outermost rim of the SNR generally conforms to our estimated CD location, which marks the boundary between the CSM and ejecta (Fig. 1c). The spectrally soft emission from the CSM may also be projected over the face of the entire SNR, ``filling in'' the regions between the spectrally-hard ejecta filaments. 

A prominent CSM feature is the central equatorial ``belt" (\citealt{park2002}, \citeyear{park2004}; \citealt{ghavamian2005}). Based on their multiband infrared $\textit{AKARI}$ observations, \citet{lee2009} proposed that this feature is actually a ring-like structure that in projection presents as two long filaments, a northern filament and a southern filament, that run east-west in the central regions of the SNR (Fig. 1d). Only the bright southern filaments of this structure were previously identified in X-rays (\citealt{park2002}). In Fig. 1c we show that the regions corresponding to the northern filament of the equatorial ring (ER), in addition to those for its southern filaments, are generally coincident with our identified \textit{CSM-dominated} regions. These filaments appear to connect into a closed loop on the eastern side but are more fragmented on the western side. The limb-brightening expected for a pure ring structure does not appear to be present in the ER (Figs. 1b--d, 5e \& f).

The ER is composed of dense gas, and both the northern and southern filaments clearly stand out in our $n_{EM}$ and $P_{EM}$ maps (Figs. 5e \& 5f). This ring-like structure may have originated from pre-SN mass loss at the equator of the rotating progenitor and/or from interactions in a binary system (e.g., \citealt{collins1999}; \citealt{chevalier2000}; \citealt{morris2009}; \citealt{langer2012}). The northern filaments of the ER are about 2-3 times fainter than the southern filaments. This may be because the ER is non-uniform, and the northern filaments have a lower gas density than the southern filaments. The non-uniform nature of the ER may also supported by the the non-detection of limb-brightening that would normally be expected for a uniform, ring-like structure. Furthermore, the northern filaments are not detected in the optical band, while the southern filaments show faint zero radial velocity [O III] emission, but no clear H$\alpha$ emission. (\citealt{ghavamian2005}). This again suggests a non-uniform structure in the ER, with the southern filaments possibly being denser than the northern filaments, thereby demonstrating radiative cooling.

Based on the ratio of the ring's N-S angular separation ($\sim$75$\arcsec$) to its E-W diameter ($\sim$295$\arcsec$), we estimate that the normal to the ring is inclined to the plane of the sky by an angle of $\alpha$ $\sim$ 15$^{\circ}$. If the ER was shed by the progenitor's equatorial winds before the SN explosion, then this inclination angle would correspond to the inclination angle of the progenitor's rotation axis. This is generally consistent with the estimated N-S alignment of the spin axis of the embedded pulsar (J1124--5916) inferred from the observed N-S orientation of its jet (\citealt{park2007}). This suggests that the rotation axis of the progenitor appears to have been preserved in the neutron star after the SN explosion.

We estimate the CSM mass in G292, by assuming a fully ionized plasma with 10\% He, giving the relationship $n_{e} = 1.2 n_{H}$. To estimate the volumes of the ER regions we assume the ER has a ring-like geometry with a line of sight depth similar to its N-S angular thickness ($\sim$15$\arcsec$). Based on this geometry we estimate the total volume of the ER $\sim$ 8\e{54} $f$ $d^{3}_{6}$ cm$^{3}$. To estimate the volume of the outer spherical CSM, we assume it has a shell-like structure with a line of sight depth similar to its angular thickness ($\sim$45$\arcsec$). For the spherical CSM projected towards the interior of the remnant, we assume a line of sight depth of 90$\arcsec$ corresponding to emission from the CSM shell both at the front, and at the back of the remnant. We calculate the total CSM mass by summing the masses for all $\sim$700 CSM regions. We list our estimates for the CSM mass in Table 5. We estimate a shocked CSM mass of 13.5$^{+1.7}_{-1.4}$ $f^{1/2}$ $d^{3/2}_{6}$ M$_{\astrosun}$. This is close to the lower bound of the previous estimates based on the radial density profile of the progenitor star's RSG winds ($\sim$15--40 M$_{\astrosun}$, \citealt{lee2010}). The ER comprises $\sim$13\% of the total estimated shocked CSM mass (Table 5). This may suggest that a significant pre-SN mass loss occurred through the progenitor's equatorial winds. The ER is reminiscent of similar ring-like circumstellar features observed in SN 1987A (e.g., \citealt{frank2016} and references therein) and the B-type supergiant SBW1 (\citealt{smith2013}). Such structures may originate from mass loss facilitated by the rapid rotation of the progenitor, binary interactions, confinement of the shed mass by magnetic fields, and possible combinations of these factors (e.g., \citealt{kurfurst2018} and references therein).  

We estimate an average thermal pressure of $\sim$5\e{-9} ergs cm$^{-3}$ for the CSM at the outer boundary of the remnant. This is comparable to the estimated ram pressure of the PWN (2.6\e{-9} $d^{-2}_6$ ergs cm$^{-3}$, \citealt{hughes2003}). This may support the interaction between the RS and PWN as suggested by \citealt{bhalerao2015}. An early-stage RS-PWN interaction scenario has also been suggested by \citet{gaensler2003}, based on the close juxtaposition between the PWN and the overlying shell in radio and X-rays. We note that previously it had been suggested that the RS may not have reached the PWN (\citealt{park2004}). This was based on the estimation of a large pressure difference between the ER and the PWN. Our pseudo thermal pressure map (Fig. 5f), shows that regions in the southern filaments of the ER generally have higher gas pressures compared to other regions in the SNR. Thus the high pressure difference estimated by \citet{park2004} may have been due to the selection of a high-pressure region from the southern filaments of the ER (Region 1 in \citealt{park2004}), rather than a region with typical gas pressure generally found between the FS and RS. 

\subsection{Ejecta Asymmetry in the Remnant}

CCSN explosions involve a complex interaction among several physical processes, the details of which are not fully understood (e.g., \citealt{janka2012}; \citealt{burrows2013}; \citealt{couch2017}). Observations indicate that a property common to many CCSNe is explosion asymmetry (e.g., \citealt{wang2001}, \citeyear{wang2003}; \citealt{leonard2006}; \citealt{wang2008}; \citealt{lopez2011}; \citealt{vink2012}; \citealt{lopez2018} and references therein). Recent theoretical studies have suggested that asymmetries in CCSNe can originate from hydrodynamic instabilities during the SN explosion (\citealt{janka2017} and references therein). These studies show that the bulk of intermediate mass elements (Si, S, Ar, Ca and Ti) and the heavier iron-group elements (Cr, Fe, Ni) are ejected opposite to the direction in which the NS is kicked. A direct correlation is also found between NS-kick velocities and ejecta asymmetries (\citealt{wongwathanarat2013}, \citeyear{wongwathanarat2017}). 

This anti-alignment or ``antipodal" asymmetry between NSs and the ejecta has been observed in several CCSNRs. Analysis of $\textit{Chandra}$ and $\textit{ROSAT}$ data with the power-ratio method showed evidence for such an asymmetry in five CCSNRs: G292, CTB 109, Cas A, PKS 1209--51 and Puppis A (\citealt{holland-ashford2017}). However, in the study by \citet{holland-ashford2017}, X-ray emission was examined in the 0.5--2.1 keV energy range (thus emissions from S, and the Fe K-shell transitions were not included), and a separation between emission originating from ejecta-rich regions as opposed to that from the CSM was not made. \citet{katsuda2018a} applied an image-decomposition method to $\textit{Chandra}$ and $\textit{XMM-Newton}$ data to find that the elements Si, S, Ar and Ca were predominantly ejected opposite to the direction of NS motion in six CCSNRs: G292, Cas A, Puppis A, Kes 73, RCW 103, and N49. High-energy X-ray studies of Cas A using \textit{NuSTAR} revealed that $^{44}$Ti, which is produced by explosive Si-burning in the inner regions of the exploding star, is distributed opposite to the direction of the NS-motion (\citealt{grefenstette2014}). $^{44}$Ti asymmetry in the form of redshifted $^{44}$Ti emission lines was also detected in SN 1987A, suggesting that its yet-undetected compact remnant may have been kicked towards the front of the remnant (\citealt{boggs2015}; \citealt{wongwathanarat2017}).

To investigate the ejecta spatial asymmetry in G292, we divided the ejecta regions into quadrants and hemispheres (Fig. 11). These divisions were made with reference to the optical expansion center of the SNR (RA (2000.) = 11$^h$, 24$^m$, 34$^s$.4 and decl. (2000.) =  -59$^{\circ}$, 15\arcmin, 51\arcsec, \citealt{winkler2009}), and the presumed direction of the pulsar's kick to the SE (\citealt{park2007}; \citealt{winkler2009}). We list the ejecta masses estimated for the quadrants and hemispheres in Table 4, and we show their fractional mass distributions in Figs. 12 \& 13. We list the average abundances for the ejecta regions in the entire remnant, as well as for the ejecta regions in the quadrants and the hemispheres in Table 6. 

The estimated ejecta masses of Si, S and Fe are significantly higher in the NW hemisphere than in the SE. The NW hemisphere accounts for $\sim$90\% of the total Si-ejecta mass, $\sim$100\% of the total S-ejecta mass, and $\sim$60\% of the total Fe-ejecta mass (Table 4, Fig. 12). A possible explanation for this non-uniform ejecta distribution may be an asymmetric RS-structure, where the RS has not propagated deep enough through the ejecta in the SE to heat the heavier Si-, S- and Fe-rich ejecta (given a layered ejecta composition). However, the  detection in the optical band, of [S II] ejecta in the southeast (\citealt{winkler2006}), does not support this explanation. Also, in the SE, the abundances of O, Ne, Mg, Fe and the ionization timescale are generally enhanced between $\sim$120$\arcsec$ and $\sim$170$\arcsec$ from the SNR center. Thus in the SE, the RS may be located at $\sim$120$\arcsec$ from the SNR center, which is similar to that in other parts of the remnant. While Fe is enhanced in the NW regions of the SNR, the overall projected distribution of the Fe-rich ejecta detection over the SNR appears to be nearly circular in shape, generally coincident with the circular outline of the RS (Fig. 3k). Thus, we propose that the apparent ``lack" of heavy elements in the SE regions of G292 is probably caused by an intrinsically asymmetric distribution of Si-, S- and Fe-rich ejecta gas rather than by a significantly asymmetric RS structure. 

Most of the Si, S and Fe ejecta are enhanced in regions opposite to the currently projected position of the pulsar (PSR J1124-5916, Figs. 1b \& 3). The pulsar is displaced $\sim$46\arcsec~from the presumed SN explosion center (RA (2000.) = 11$^h$, 24$^m$, 34$^s$.4 and decl. (2000.) =  -59$^{\circ}$, 15\arcmin, 51\arcsec, \citealt{winkler2009}). This displacement corresponds to an average transverse velocity of $\sim$440 km s$^{-1}$ at $d$ = 6 kpc and age $\sim$ 3000 yr (\citealt{hughes2001}, \citeyear{hughes2003}; \citealt{winkler2009}). This suggested motion of the pulsar is likely the result of a ``kick" to the newly-formed NS during the SN explosion (\citealt{lai2001}; \citealt{park2007}; \citealt{winkler2009}). Recent 2D and 3D numerical simulations demonstrate that significant NS-kicks in the range of a few hundred to $\sim$1000 km s$^{-1}$ can be produced by asymmetric SN explosions in which the bulk of intermediate and iron-group elements (e.g., Si and Fe) are ejected in the opposite direction to the NS-kick (e.g., \citealt{wongwathanarat2013}, \citeyear{wongwathanarat2017}; \citealt{janka2017} and references therein). Consequently, the NS and the bulk of the inner ejecta (the explosive nucleosynthesis products) would move in opposite directions of each other in the resulting SNR. The kick imparted to the NS mainly originates from gravitational and hydrodynamic forces exerted by the asymmetric ejecta on the NS, and from momentum conservation. Anisotropic neutrino emission plays only a minor role (\citealt{scheck2006}; \citealt{wongwathanarat2013}; \citealt{janka2017}). The NS is accelerated by the gravitational forces exerted by the slower, denser ejecta found in the hemisphere opposite to the stronger explosion. The acceleration of the NS can last several seconds, so that while the NS is pulled into one hemisphere, explosive nucleosynthesis of Fe-group elements proliferates in the opposite hemisphere (\citealt{scheck2006}; \citealt{nordhaus2010}, \citeyear{nordhaus2012}; \citealt{wongwathanarat2013}, \citeyear{wongwathanarat2017}; \citealt{janka2017}).

In this study, we provide the first observational evidence that the bulk of Fe was ejected opposite to the NS's kick (e.g., Figs. 3k, 3n, 4f, 9a \& 12).  This spatial distribution of the Fe-rich ejecta is in close agreement with 3D simulations which predict that $^{56}$Ni, the parent-isotope of $^{56}$Fe, is mainly formed and ejected opposite to the NS's kick vector (\citealt{wongwathanarat2013}, \citeyear{wongwathanarat2017}). Along with Fe, we also show that the bulk of Si and S was ejected opposite to the NS's kick (e.g., Figs. 3, 4 \& 12). Considering the concentration of dense, cooler ejecta gas in the SE quadrant of the SNR, as observed in the optical and infrared bands (\citealt{ghavamian2005}, \citeyear{ghavamian2009}, \citeyear{ghavamian2012}; \citealt{winkler2006}; \citealt{winkler2009}), it is plausible that asymmetries in the explosion led to density variations in the ejecta. As a result, the O-, and Ne-rich ejecta expelled to the southeast are denser and clumpier than they are in other parts of the remnant. Thus, as suggested by numerical simulations (e.g., \citealt{wongwathanarat2013}, \citeyear{wongwathanarat2017}; \citealt{janka2017}), the NS may also have been gravitationally pulled to the southeast by these denser, slower moving, O-, Ne-rich ejecta. These ejecta are now radiatively cooling, and appear as the prominent crescent-shaped spur observed in the optical and infrared bands (\citealt{ghavamian2005}, \citeyear{ghavamian2009}, \citeyear{ghavamian2012}; \citealt{winkler2006}; \citealt{winkler2009}). 

Our results suggest an explosion axis generally oriented NW-SE. We find that the X-ray-emitting O-, Ne- and Mg-rich ejecta are generally enhanced along the NW-SE direction (Fig. 3). To test the explosion kinematics projected across the plane of the sky (as suggested by this ejecta mass distribution), we estimated the O-, Ne- and Mg-rich ejecta mass along the main shell of the SNR (regions between our estimated RS and CD, Fig. 13). We excluded regions projected within the RS for this comparison, since these ejecta may have their main kinematics along the line of sight. The combined NW+SE shell-regions account for a significantly higher fraction of the ejecta mass for O, Ne and Mg ($\sim$60\%--65\%, Fig. 13 and Table 7).

\subsection{Progenitor Mass}
To estimate the progenitor mass for G292, we compared the elemental abundance ratios for ejecta regions from the entire SNR to those predicted by CCSN nucleosynthesis models (\citealt{woosley1995}). In Fig. 14 we show those modeled values of elemental abundance ratios for which the progenitor masses are relatively well-discriminated. We compare these modeled ratios to our estimated mean abundance ratios for G292. Based on these comparisons, and taking into account all five plots in Fig. 14, we place an upper limit to the G292 progenitor mass of $\lesssim$ 30 M$_{\astrosun}$. 

From our measured Si/Fe ratio for the \textit{ejecta-dominated} regions in G292, we obtain a progenitor mass estimate of $\sim$13--15 M$_{\astrosun}$ (Fig. 14b). Our measured O/Fe ratio for the \textit{ejecta-dominated} regions in G292 corresponds to a progenitor mass of $\sim$15-25 M$_{\astrosun}$ (Fig. 14a). These two ratios suggest a lower limit to the G292 progenitor mass of $\gtrsim$ 13 M$_{\astrosun}$. Recently, \citet{katsuda2018b} propose that both of these ratios (Fe/O and Fe/Si) are more reliable indicators of the CCSN progenitor mass than the traditionally used abundance ratios of the other alpha elements to Si (e.g., O/Si, Ne/Si, Mg/Si, and S/Si). This is based on their comparisons of the progenitor mass-distributions of 41 CCSNRs in the Milky Way and the Large and Small Magellanic Clouds, to the standard Salpeter initial mass function. Based on G292's Fe/Si ratio measured previously with \textit{Suzaku} data (\citealt{kamitsukasa2014}), they estimated a progenitor mass of  $\lesssim$15 M$_{\astrosun}$. Our estimated lower limit is consistent with this estimate by \citet{katsuda2018b}. Thus, our abundance ratios suggest a progenitor mass in the range of 13 M$_{\astrosun}$ $\lesssim$ M $\lesssim$ 30 M$_{\astrosun}$ (Fig. 14). 

If we combine our calculated mass for the shocked ejecta (1.29$^{+0.49}_{-0.28}$ $f^{1/2}$ $d^{3/2}_{6}$ M$_{\astrosun}$, Table 4), the shocked CSM (13.5$^{+1.7}_{-1.4}$ $f^{1/2}$ $d^{3/2}_{6}$ M$_{\astrosun}$, Table 5) with a typical neutron star mass of 1.4 M$_{\astrosun}$ and a dust mass of 0.023 M$_{\astrosun}$ estimated for G292 (\citealt{ghavamian2016}), we obtain total mass of $\sim$16$^{+2.2}_{-1.7}$ M$_{\astrosun}$. This mass is close to the lower bound of our estimated progenitor mass range based on the ejecta elemental abundance ratios. The progenitor mass would be higher than this calculated value depending on the mass of the unshocked ejecta, and the unshocked CSM mass lost prior to the explosion through stellar winds and binary interactions. Optically-emitting material would also add to the progenitor mass. However, the mass of the optical material may be low ($\lesssim$ a few solar masses) given that the optical emission in G292 (e.g., the integrated [O III] flux) is estimated to be about 20$\times$ less than the X-ray flux (\citealt{winkler2006}). It is not clear if the bulk of the O, Ne and Mg ejecta have already been shocked in G292. Turbulence and mixing may have retained ejecta for these lighter elements close to the center of the explosion. There could also be reserves of unshocked Si-, S- and Fe-rich ejecta close to the SNR's center (see Section 4.2). We estimate that there could be the following amounts of unshocked ejecta in G292: O $\lesssim$ 2.8 M$_\odot$, Ne $\lesssim$ 0.4 M$_\odot$, Mg $\lesssim$ 0.15 M$_\odot$, Si $\lesssim$ 0.2 M$_\odot$, S $\lesssim$ 0.1 M$_\odot$, and Fe $\lesssim$ 0.07 M$_\odot$ (based on yields calculated for progenitors with mass 13--30 M$_\odot$, $Z$ = 0.02 and $E$ = 1\e{51} ergs, \citealt{nomoto2006}). Thus there could be as much as 4 M$_{\astrosun}$ of ejecta that have not yet been shocked by the RS in G292. We note that analyses of G292's expansion dynamics have suggested similar total ejecta mass values. Analysis of G292's dynamics in the radio band has suggested a total ejecta mass of $\sim$5.9 M$_{\astrosun}$ (\citealt{gaensler2003}), and analysis of the remnant's dynamics in the X-ray band has suggested a total ejecta mass of $\lesssim$ 8 M$_{\astrosun}$ (\citealt{bhalerao2015}).

We compared our measured elemental ejecta-mass yields to those predicted for CCSN nucleosynthesis models (\citealt{nomoto2006}, for $Z$ = 0.02 and $E$ = 1\e{51} ergs, Fig. 15). Since our estimated total ejecta mass is only the shocked component (and thus represents a lower limit of the total ejecta mass), this comparison allows us to rule out progenitor masses with elemental ejecta yields less than our estimated values. The yields of O, Ne and Mg suggest a progenitor with a mass $\gtrsim$ 20 M$_\odot$, with a lower limit of $\gtrsim$ 15 M$_{\odot}$ (Fig. 15). We note that predicted yields are model-dependent, and are sensitive to several parameters such as metallicity and rotation rate of the progenitor, and specifics of the explosion (e.g., \citealt{hirschi2017}; \citealt{fryer2018}). We also note that our ejecta and CSM mass estimates may involve systematic uncertainties associated with our assumed X-ray emission volumes for individual emission regions, and the clumpiness of the emission (which is embedded in the volume filling factor $f$).

Theoretical calculations generally favor the production of a black hole rather than a NS for a progenitor of mass $\gtrsim$ 25 M$_{\astrosun}$ (e.g., \citealt{woosley2002}). Exceptions to this ``traditional" limit may exist; for example, recent CCSN explosion models indicate that black holes can result from progenitor explosions of M $\gtrsim$ 15 M$_{\astrosun}$, and NSs can form from progenitors as massive as 120 M$_{\astrosun}$ (\citealt{sukhbold2016}). Various factors such as the progenitor's metallicity, the presence of stellar companions, rotation rate of the progenitor, explosion energy and nucleosynthesis history can affect the outcome of the explosion. For example, a significant mass loss by a massive progenitor (with M $\gtrsim$ 25 M$_{\astrosun}$) to its companion in a binary system, may lead to the formation of a NS rather than a black hole (e.g., \citealt{belczynski2008} and references therein). The presence of the ER in G292 suggests that its progenitor star may have been part of a binary and/or rapidly rotating (see Section 4.3). Thus if the G292 progenitor was more massive than $\sim$20--25 M$_{\odot}$, the formation of the observed pulsar rather than a black hole may be the result of significant mass loss of the progenitor star due to its binary interaction prior to the SN explosion. If the binary companion survived the explosion, it may be located within a radius of $\sim$70\arcsec\,from the explosion site, assuming a high runaway velocity of $\sim$500 km s$^{-1}$ (e.g., \citealt{eldridge2011}), and a distance to the remnant of 6 kpc. Possible contamination of the companion star's spectrum by metal-rich ejecta from G292 may also help in the search for the surviving companion (e.g., \citealt{gvaramadze2017}; \citealt{hirai2018}; \citealt{liu2018}). From the above results of our elemental abundance ratio comparisons and ejecta yield estimates, we propose a conservative estimate of 13 M$_{\odot}$ $\lesssim$ M $\lesssim$ 30 M$_{\odot}$ for the G292 progenitor mass. A massive progenitor (M $\gtrsim$ 15 M$_{\odot}$), would imply an O or B main sequence star. Such a star would have produced an extended H II region around itself. Some of the diffuse, structured, optical H$\alpha$ and [S II] emission observed around G292 (\citealt{winkler2006}), is probably a relic of this H II region.

\section{\label{sec:summ} SUMMARY \& CONCLUSIONS}

Based on our deep \textit{Chandra data}, we study the detailed structure of the shocked ejecta and CSM gas in the textbook-type supernova remnant G292.0+1.8. Our results are based on the systematic spectral analysis of over $\sim$2000 regions covering the entire remnant. We identify the spatial distribution of Fe-rich ejecta in G292 for the first time, based on our detection of enhanced Fe L and Fe K-shell line emission. We provide spatial distribution maps for O-, Ne-, Mg- Si-, S- and Fe-rich ejecta. For the first time we identify the X-ray counterparts of the entire equatorial ring-like dense CSM, whose components had previously been detected in infrared. Based on elemental abundance ratios, and estimates of the ejecta and CSM masses, we estimate the G292 progenitor star's main sequence mass of 13--30 M$_{\astrosun}$. Our ejecta maps reveal a preferred NW-SE distribution axis for the elements O, Ne, Mg, Si, S and Fe. Furthermore, Si-, S- and Fe-rich ejecta are primarily found in the northwestern hemisphere. This provides the first clear observational evidence that the inner ejecta in G292, consisting of key explosive nucleosynthesis products such as Fe, were predominantly expelled opposite to the direction of the neutron star's kick during the SN explosion. This anti-alignment between the neutron star and the heavier inner ejecta is consistent with theoretical CCSN calculations in which the neutron star-kick originates from gravitational and hydrodynamic forces exerted by the asymmetric ejecta on the NS, and from momentum conservation.

\acknowledgments

This work was supported in part by the SAO through \textit{Chandra} grants GO1-12077X and GO6-7049A. JPH acknowledges support for X-ray studies of supernova remnants from NASA grant NNX15AK71G. JB acknowledges support from the NASA Texas Space Grant Consortium. We thank Dr. Daniel Dewey, and Dr. Thomas G. Pannuti for valuable discussions. We also thank the MIT \textit{Chandra} HETG group for useful ISIS scripts and plotting routines.
\software{CIAO\, (\citealt{fruscione2006}),\, XSPEC\, (\citealt{arnaud1996}),\, ISIS\, (\citealt{houck2000}).}


\begin{deluxetable*}{lccc}[b!]
\tablecaption{$\textit{Chandra}$ Observations of G292 \label{tab:ObsIDs}}
\tablecolumns{4}
\label{tbl:tab1}
\tablewidth{0pt}
\tablehead{
\colhead{ObsID} & \colhead{Date} & \colhead{Exposure} & \colhead{Roll Angle\tablenotemark{a}} \\
\colhead{} & \colhead{} & \colhead{(ks)} & \colhead{($^{\circ}$)} }
\startdata
6680\tablenotemark{b} & 2006 Sep 13 & 39 & 180 \\
6678 & 2006 Oct 2 & 44 & 157  \\
6679 & 2006 Oct 3 & 154 & 157  \\
8447 & 2006 Oct 7 & 48& 157  \\
6677 & 2006 Oct 16 & 159 & 140  \\
8221 & 2006 Oct 20 & 65& 140  \\
\enddata
\tablenotetext{a}{The roll angle describes the orientation of the $\textit{Chandra}$ Observatory in the sky, as it is rotated about the viewing axis to optimally align its solar panels with the Sun.}
\tablenotetext{b}{ObsID 6680 included two ACIS-S chips in addition to four ACIS-I chips. This observation was affected by telemetry saturation which reduced the exposure time by $\sim$25$\%$. The 5 subsequent observations used the four ACIS-I chips only, and were not affected by telemetry saturation. } 

\end{deluxetable*}

\begin{deluxetable*}{cccccccccc}
\footnotesize
\tabletypesize{\scriptsize}
\setlength{\tabcolsep}{0.05in}
\tablecaption{Average CSM Abundances}
\label{tbl:tab2}
\tablewidth{0pt}
\tablehead{\\ \colhead{$N_{\rm H}$} & \colhead{$kT$} & \colhead{$n_{e}t$} & \colhead{\it EM} & \colhead{O} & \colhead{Ne} & \colhead{Mg} & \colhead{Si} & \colhead{S} & \colhead{Fe} \\
\colhead{(10$^{22}$~cm$^{-2}$)} & \colhead{(keV)} & \colhead{(10$^{12}$ cm$^{-3}$ s)} & \colhead{($10^{56}$ cm$^{-3}$)} & \colhead{} & \colhead{} & \colhead{} & \colhead{} & \colhead{} & \colhead{} }
\startdata
$0.43^{+0.02}_{-0.02}$ & $0.82^{+0.14}_{-0.03}$ & $0.24^{+0.05}_{-0.03}$ & $1.37^{+0.13}_{-0.16}$ &  $0.33^{+0.06}_{-0.03}$ & $0.48^{+0.05}_{-0.03}$ & $0.24^{+0.03}_{-0.02}$ & $0.21^{+0.03}_{-0.02}$ & $0.34^{+0.08}_{-0.08}$  & $0.13^{+0.03}_{-0.02}$ \enddata
\tablecomments{Abundances are with respect to solar (\citealt{anders1989}). Uncertainties are at the 90\% confidence level.}
\end{deluxetable*}

\begin{deluxetable*}{lccc}
\footnotesize
\tabletypesize{\scriptsize}
\setlength{\tabcolsep}{0.05in}
\tablecaption{Line and Continuum Energy Ranges for the EWI Images}
\label{tbl:tab3}
\tablewidth{0pt}
\tablehead{\\ \colhead{Element/Ion} & \colhead{Line} & \colhead{Low\tablenotemark{a} } & \colhead{High\tablenotemark{a}} \\
\colhead{} & \colhead{(eV)} & \colhead{(eV)} & \colhead{(eV)} }
\startdata
O He$\alpha$~\&~Ly$\alpha$ (O VII~\&~O VIII)& $510-740$ & $300-510$ & $740-870$ \\
Ne He$\alpha$ (Ne IX) & $890-970$ & $740-870$ & $1120-1160$ \\
Ne Ly$\alpha$ (Ne X) & $1000-1100$ & $740-870$ & $1120-1160$ \\
Mg He$\alpha$ (Mg XI)& $1290-1420$ & $1250-1290$ & $1620-1700$ \\
Mg Ly$\alpha$ (Mg XII) & $1430-1510$ & $1390-1430$ & $1520-1630$ \\
Si He$\alpha$ (Si XIII) & $1750-1930$  & $1620-1700$ & $2020-2120$ \\
S He$\alpha$ (S XV) & $2300-2600$  & $2000-2100$ & $2610-2750$ \\
Fe He$\alpha$ (Fe XXV) & $6300-6850$ & $5500-5900$ & $6950-7250$
\enddata
\tablenotetext{a}{Low and high energy ranges used to estimate the underlying continuum.}
\end{deluxetable*}

\begin{deluxetable*}{lccccccc}
\footnotesize
\tabletypesize{\scriptsize}
\setlength{\tabcolsep}{0.05in}
\tablecaption{Estimated Shocked Ejecta Masses}
\label{tbl:tab4}
\tablewidth{0pt}
\tablehead{\\ \colhead{Element} & \colhead{Entire SNR} & \colhead{NW} & \colhead{NE} & \colhead{SE} & \colhead{SW} & \colhead{NW Hem.\tablenotemark{*}} & \colhead{SE Hem.\tablenotemark{*}} \\
\colhead{} & \colhead{(M$_{\astrosun}$)} & \colhead{(M$_{\astrosun}$)} & \colhead{(M$_{\astrosun}$)} & \colhead{(M$_{\astrosun}$)} & \colhead{(M$_{\astrosun}$)} & \colhead{(M$_{\astrosun}$)} & \colhead{(M$_{\astrosun}$)}  } 
\startdata
O & $0.47^{+0.23}_{-0.12}$ & $0.13^{+0.06}_{-0.03}$ & $0.09^{+0.03}_{-0.02}$ & $0.08^{+0.05}_{-0.03}$  & $0.08^{+0.05}_{-0.03}$ & $0.20^{+0.09}_{-0.05}$ & $0.17^{+0.10}_{-0.05}$\\
Ne & $0.57^{+0.18}_{-0.10}$ & $0.16^{+0.04}_{-0.03}$ & $0.10^{+0.02}_{-0.01}$ & $0.08^{+0.03}_{-0.02}$ & $0.11^{+0.05}_{-0.03}$ & $0.27^{+0.07}_{-0.04}$ & $0.19^{+0.08}_{-0.04}$\\
Mg & $0.13^{+0.04}_{-0.03}$ & $0.06^{+0.02}_{-0.01}$ & $0.04^{+0.01}_{-0.01}$ & $0.03^{+0.01}_{-0.01}$ & $0.04^{+0.02}_{-0.01}$ & $0.10^{+0.02}_{-0.02}$ & $0.07^{+0.03}_{-0.02}$\\
Si & $0.06^{+0.01}_{-0.01}$ & $0.01^{+0.003}_{-0.002}$ & $0.008^{+0.002}_{-0.001}$ & $0.001^{+0.0001}_{-0.0001}$ & $0.005^{+0.002}_{-0.001}$ & $0.02^{+0.005}_{-0.004}$ & $0.002^{+0.0006}_{-0.0006}$\\
S & $0.02^{+0.01}_{-0.01}$ & $0.003^{+0.001}_{-0.001}$ & $0.002^{+0.0003}_{-0.0003}$ & $0.0$$^\tablenotemark{\textdagger}$ & $0.0003^{+0.0001}_{-0.0001}$ & $0.005^{+0.001}_{-0.001}$ & $0.0$$^\tablenotemark{\textdagger}$\\
Fe & $0.03^{+0.01}_{-0.01}$ & $0.008^{+0.003}_{-0.001}$ & $0.005^{+0.001}_{-0.001}$ & $0.003^{+0.002}_{-0.001}$ & $0.005^{+0.003}_{-0.002}$ & $0.01^{+0.004}_{-0.002}$ & $0.008^{+0.004}_{-0.002}$\\
Total & $1.29^{+0.49}_{-0.28}$ & $0.37^{+0.13}_{-0.07}$ & $0.24^{+0.06}_{-0.04}$ & $0.19^{+0.10}_{-0.06}$ & $0.25^{+0.12}_{-0.06}$ & $0.61^{+0.20}_{-0.11}$ & $0.44^{+0.21}_{-0.12}$ \enddata
\tablecomments{Ejecta masses are in terms of $f^{1/2}d^{3/2}_{6} M_{\astrosun}$, where $d_{6}$ is the distance to the SNR in units of 6 kpc. The mass for the entire SNR is the sum of the six elemental masses calculated for all $\sim$1400 ejecta regions. The masses for the NW, NE, SE and SW quadrants and the NW and SE hemispheres are for regions where the given element's abundance was higher than the CSM abundance by at least a 3$\sigma$ level of confidence. We note that there may be additional amounts of radiatively cooled ejecta that are emitting predominantly at optical and infrared wavelengths. For example, significant ejecta emission from [O III], [N II], and [N IV] has been detected in a crescent-shaped structure in the remnant's southeast known as the ``spur'' (e.g., \citealt{ghavamian2005}, \citeyear{ghavamian2009}, \citeyear{ghavamian2012}; \citealt{winkler2006}; \citealt{winkler2009}). Thus, the total ejecta mass in G292 may be higher than our estimates here, that only consider the X-ray-emitting ejecta.
\tablenotetext{*}{Hem.: Hemisphere}
\tablenotetext{$\textdagger$}{We note that, in the optical band, a few S-rich ejecta features have been detected in the southeastern parts of the remnant (\citealt{winkler2006}). This suggests that there may be dense clumps of S-rich ejecta in  the southeast that have radiatively cooled to produce optical emission but not X-rays.}}

\end{deluxetable*}

\begin{deluxetable*}{cc}
\footnotesize
\tabletypesize{\scriptsize}
\setlength{\tabcolsep}{0.05in}
\tablecaption{Estimated Shocked CSM Masses}
\label{tbl:tab5}
\tablewidth{0pt}
\tablehead{\\ \colhead{Component} & \colhead{Mass (M$_{\astrosun}$)} } 
\startdata
Equatorial ring & $1.7^{+0.1}_{-0.1}$ \\
Outer spherical CSM & $5.0^{+0.6}_{-0.6}$ \\
Centrally projected spherical CSM & $6.8^{+1.0}_{-0.7}$ \\
Total & $13.5^{+1.7}_{-1.4}$   \\
\enddata
\tablecomments{The masses are in terms of $f^{1/2}d^{3/2}_{6} M_{\astrosun}$, where $d_{6}$ is the distance to the SNR in units of 6 kpc. The ``spherical CSM" includes all the regions with spectrally soft, diffuse emission. This includes the ``outer spherical CSM" regions which are projected close to the outer boundary of the remnant, and the ``centrally projected spherical CSM" regions which are projected towards the central parts of the remnant.}
\end{deluxetable*}

\begin{deluxetable*}{lcccccc}
\footnotesize
\tabletypesize{\scriptsize}
\setlength{\tabcolsep}{0.05in}
\tablecaption{Average Ejecta Abundances}
\label{tbl:tab6}
\tablewidth{0pt}
\tablehead{\\ \colhead{Region} & \colhead{O} & \colhead{Ne} & \colhead{Mg} & \colhead{Si} & \colhead{S} & \colhead{Fe}}
\startdata
Entire SNR: & $1.63^{+0.05}_{-0.02}$ & $2.63^{+0.03}_{-0.01}$ & $1.24^{+0.01}_{-0.01}$ & $0.84^{+0.01}_{-0.01}$ & $1.55^{+0.06}_{-0.04}$  & $1.26^{+0.07}_{-0.02}$\\
NW quadrant: & $1.82^{+0.09}_{-0.03}$ & $3.18^{+0.06}_{-0.02}$ & $1.47^{+0.03}_{-0.01}$ & $0.89^{+0.02}_{-0.01}$ & $1.74^{+0.10}_{-0.08}$  & $1.51^{+0.15}_{-0.03}$\\
NE quadrant: & $1.43^{+0.05}_{-0.02}$ & $2.44^{+0.04}_{-0.02}$ & $1.12^{+0.02}_{-0.01}$ & $0.96^{+0.03}_{-0.02}$ & $1.64^{+0.10}_{-0.08}$  & $1.54^{+0.08}_{-0.02}$ \\
SE quadrant: & $1.75^{+0.17}_{-0.05}$ & $2.38^{+0.14}_{-0.04}$ & $1.20^{+0.05}_{-0.02}$ & $0.65^{+0.11}_{-0.09}$ & $1.35^{+1.08}_{-0.74}$  & $0.83^{+0.21}_{-0.06}$ \\
SW quadrant: & $1.51^{+0.09}_{-0.04}$ & $2.22^{+0.05}_{-0.02}$ & $1.08^{+0.02}_{-0.01}$ & $0.68^{+0.02}_{-0.01}$ & $1.16^{+0.08}_{-0.07}$  & $0.78^{+0.09}_{-0.06}$ \\
NW Hemisphere: & $1.68^{+0.06}_{-0.02}$ & $2.76^{+0.04}_{-0.01}$ & $1.29^{+0.02}_{-0.01}$ & $0.87^{+0.02}_{-0.01}$ & $1.63^{+0.06}_{-0.05}$  & $1.63^{+0.11}_{-0.02}$ \\
SE Hemisphere: & $1.56^{+0.08}_{-0.03}$ & $2.42^{+0.06}_{-0.02}$ & $1.16^{+0.02}_{-0.01}$ & $0.65^{+0.03}_{-0.03}$ & $0.92^{+0.13}_{-0.12}$  & $0.82^{+0.07}_{-0.03}$ \\
\enddata

\tablecomments{Abundances are with respect to solar (\citealt{anders1989}). Uncertainties are averages of the 90\% confidence level uncertainties. The locations of the quadrants and hemispheres are shown in Fig. 11.}
\end{deluxetable*}

\begin{deluxetable*}{lcccc}
\footnotesize
\tabletypesize{\scriptsize}
\setlength{\tabcolsep}{0.05in}
\tablecaption{Estimated Shocked Ejecta Masses - Shell Regions}
\label{tbl:tab7}
\tablewidth{0pt}
\tablehead{\\ \colhead{Element} & \colhead{NW} & \colhead{NE} & \colhead{SE} & \colhead{SW}  \\
\colhead{} & \colhead{(M$_{\astrosun}$)} & \colhead{(M$_{\astrosun}$)} & \colhead{(M$_{\astrosun}$)} & \colhead{(M$_{\astrosun}$)}  } 
\startdata
O & $0.10^{+0.05}_{-0.03}$ & $0.06^{+0.02}_{-0.01}$ & $0.04^{+0.02}_{-0.01}$ & $0.02^{+0.01}_{-0.01}$  \\
Ne & $0.13^{+0.03}_{-0.02}$ & $0.07^{+0.02}_{-0.01}$ & $0.05^{+0.02}_{-0.01}$ & $0.05^{+0.01}_{-0.01}$  \\
Mg & $0.05^{+0.01}_{-0.01}$ & $0.02^{+0.005}_{-0.003}$ & $0.02^{+0.006}_{-0.004}$ & $0.02^{+0.004}_{-0.003}$  \\
Si & $0.006^{+0.002}_{-0.001}$ & $0.006^{+0.001}_{-0.001}$ & $0.0004^{+0.0001}_{-0.0001}$ & $0.005^{+0.001}_{-0.001}$  \\
S & $0.002^{+0.001}_{-0.001}$ & $0.001^{+0.0002}_{-0.0002}$ & $0.0$ & $0.0003^{+0.0001}_{-0.0001}$ \\
Fe & $0.005^{+0.002}_{-0.001}$ & $0.003^{+0.001}_{-0.001}$ & $0.002^{+0.0003}_{-0.0004}$ & $0.002^{+0.001}_{-0.001}$  \\
Total & $0.29^{+0.10}_{-0.06}$ & $0.16^{+0.04}_{-0.02}$ & $0.11^{+0.05}_{-0.03}$ & $0.10^{+0.03}_{-0.02}$ \enddata
\tablecomments{Ejecta masses are in terms of $f^{1/2}d^{3/2}_{6} M_{\astrosun}$, where $d_{6}$ is the distance to the SNR in units of 6 kpc. The masses are calculated for ejecta regions where the given element's abundance was higher than the CSM abundance by at least a 3$\sigma$ level of confidence. Only regions projected between the estimated positions of the RS and CD (\citealt{bhalerao2015}) are included in this table. These ``shell'' regions are shown in Fig. 13a.}
\end{deluxetable*}

\begin{figure*}[]
\figurenum{1}
\begin{center}
\includegraphics[angle=0,scale=0.18]{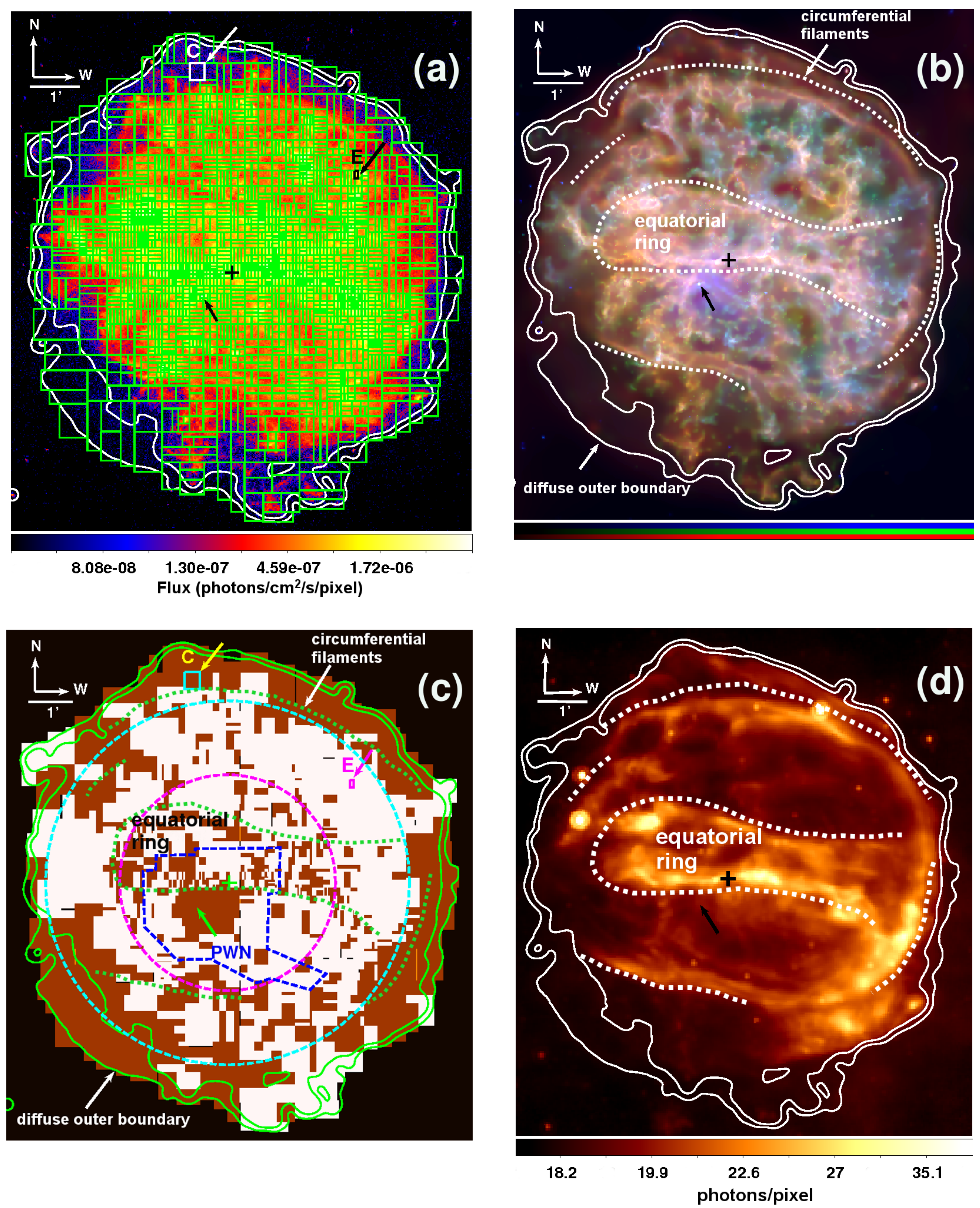}\\
\end{center}
\figcaption[]{{\textbf{(a)} ACIS-I broadband image of G292 (0.3-8 keV) overlaid with the 2147 regions (green boxes) made with our adaptive mesh method. Each region contains $\sim$5000 counts. \textbf{(b)} ACIS-I three-color image of G292 (red = 0.3-0.8 keV, green = 0.8-1.7 keV and blue = 1.7-8.0 keV). \textbf{(c)} Ejecta and CSM map constructed based on measured abundances for O, Ne, Mg, Si, S and Fe. Ejecta-dominated regions are white, while CSM dominated regions are brown. The overlaid dashed circles are our estimated locations of the reverse shock (RS, magenta) and contact discontinuity (CD, cyan), whose positions were inferred from radial velocity measurements of ejecta knots (\citealt{bhalerao2015}). The blue-dashed line marks the location of the 310 PWN regions. \textbf{(d)} 24 $\mu$m \textit{Spitzer} image of G292 (\citealt{ghavamian2012}). In all four images, the outermost contours (green in {\it a}, {\it c} and white in {\it b}, {\it d}) mark the outer boundary of the SNR in the X-ray band (E = 0.3-8 keV). The cross at the center marks the optical expansion center (\citealt{winkler2009}), and the arrow southeast of the center marks the position of the pulsar (PSR J1124-5916). The dashed curves in {\it b}, {\it c} and {\it d} highlight the locations of the thin circumferential filaments and equatorial ring which were previously identified as candidate CSM regions in X-ray (\citealt{gonzalez2003}; \citealt{park2002}, \citeyear{park2007}) and infrared studies (\citealt{lee2009}; \citealt{ghavamian2012}). The regions labeled ``C" and ``E" in {\it a} and {\it c}, are sample CSM-dominated and ejecta-dominated regions respectively, whose spectra are shown in Fig. 2.}}\label{fig:Figure 1}
\end{figure*}

\begin{figure*}[]
\figurenum{2}
\begin{center}
\includegraphics[angle=0,scale=0.30]{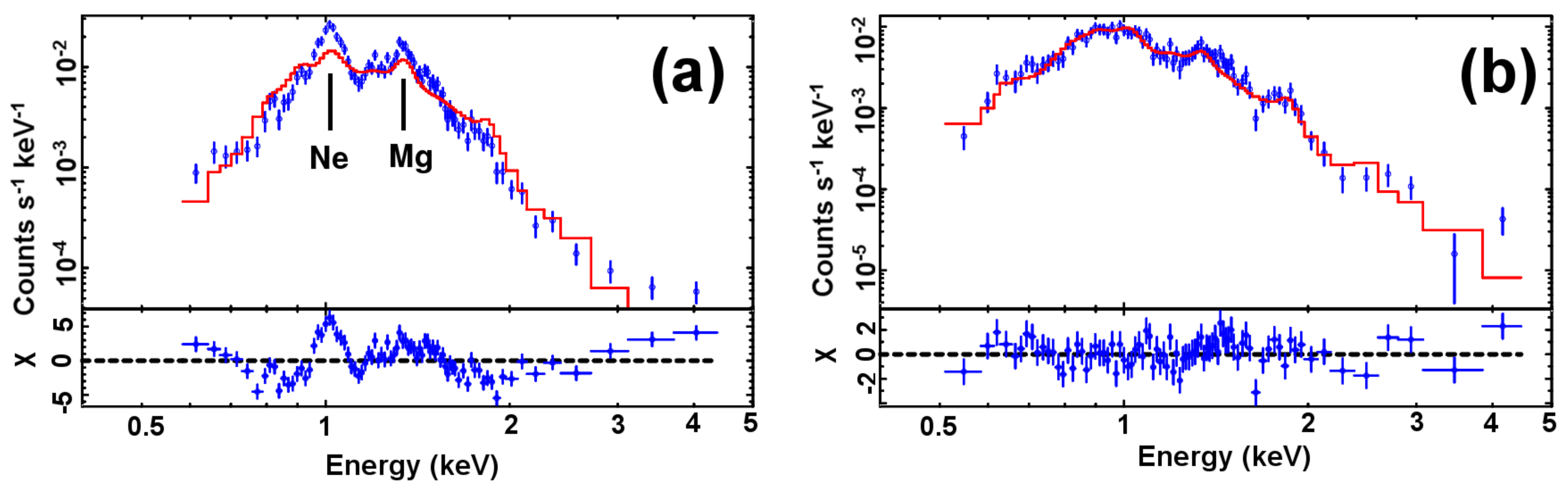}\\
\end{center}
\figcaption[]{{\textbf{(a)} An example spectrum of an ejecta region (region ``E" marked in Figs. 1a \& c), showing strong emission lines for Ne and Mg. The red line is the fitted plane shock model assuming CSM abundances. The large residuals (bottom panel) indicate a poor fit resulting in a high $\chi^2/\nu$ value of 6.6 for this region. \textbf{(b)} An example spectrum of a CSM-dominated region near the outer boundary (region ``C" marked in Figs. 1a \& c) showing low abundances and thus giving a good fit ($\chi^2/\nu$ = 1.2) to the CSM model.}\label{fig:Figure 1b}}
\end{figure*}

\begin{figure*}[]
\figurenum{3}
\begin{center}
\includegraphics[angle=0,scale=0.25]{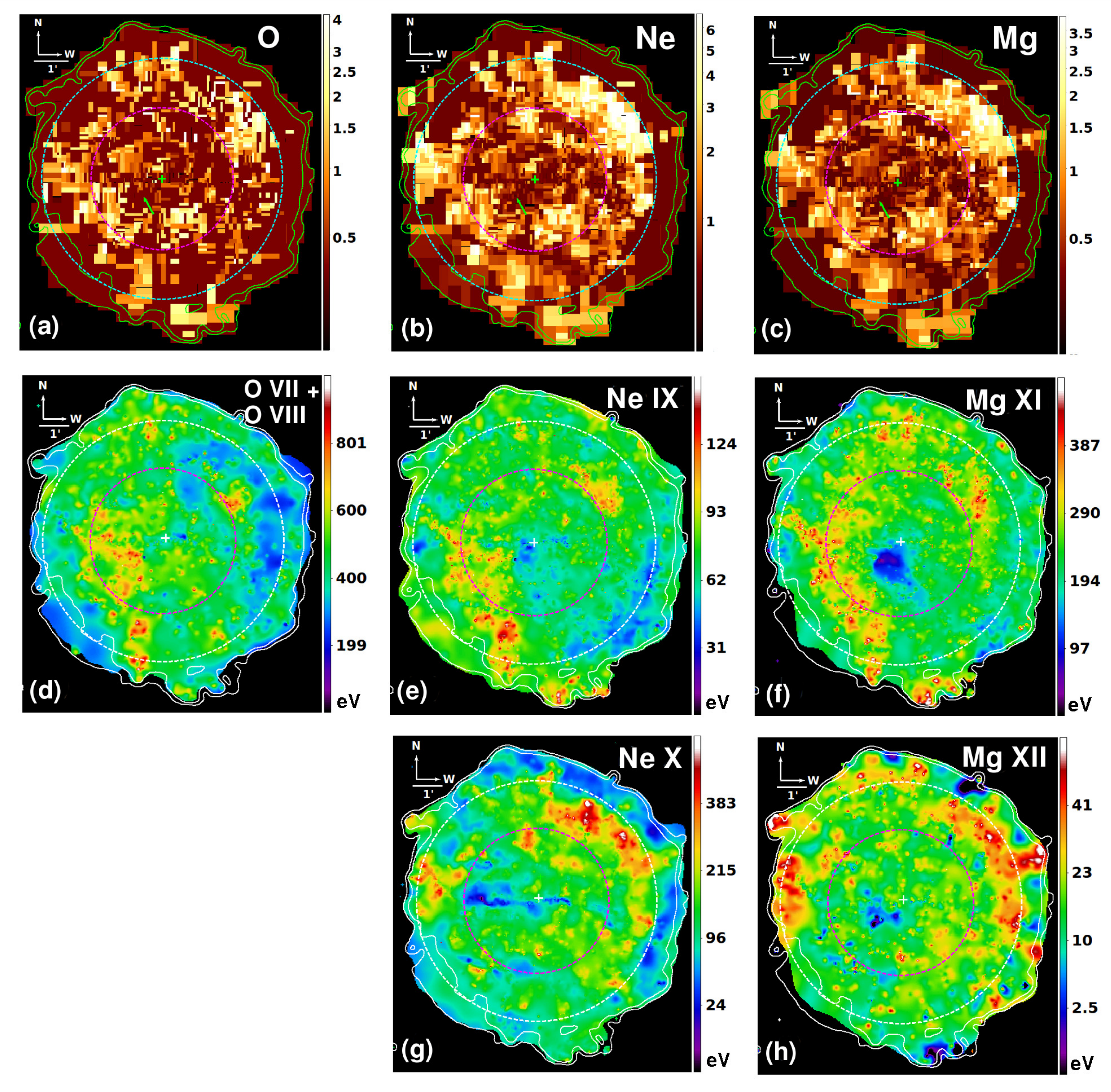}\\
\end{center}
\figcaption[]{{}\label{fig:Figure 3}}
\end{figure*}

\begin{figure*}[]
\figurenum{3}
\begin{center}
\includegraphics[angle=0,scale=0.25]{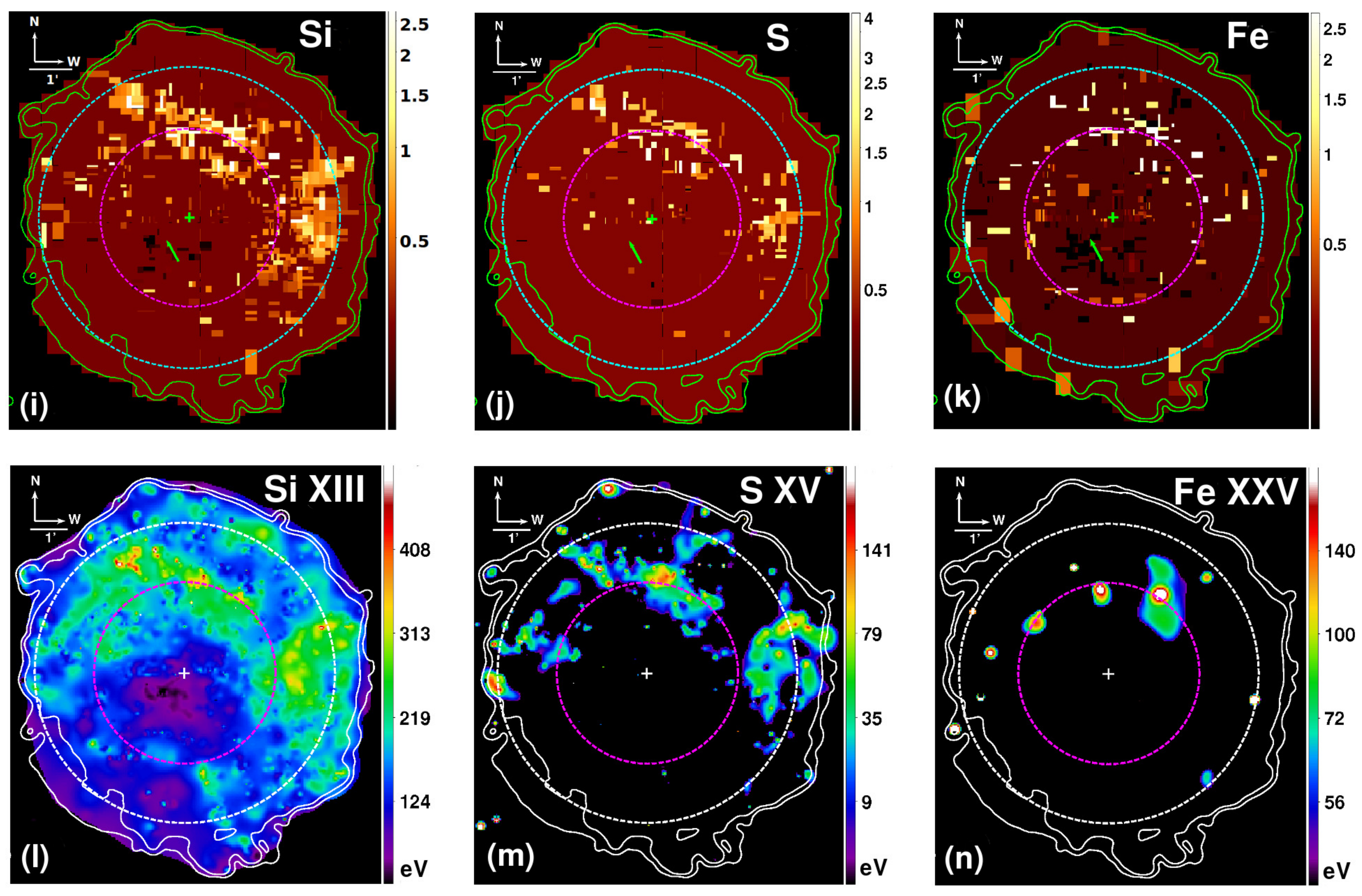}\\
\end{center}
\figcaption[]{{\textbf{Top panels (a, b, c, i, j \& k):} O, Ne, Mg, Si, S and Fe abundances (in units of solar, \citealt{anders1989}). The magenta and cyan dashed-circles represent the estimated locations of the RS and CD respectively (\citealt{bhalerao2015}). The green contours mark the outer boundary of the SNR in the X-rays based on the 0.3-8 keV band ACIS-I image. The cross marks the optical expansion center (\citealt{winkler2009}), and the arrow marks the pulsar (PSR J1124-5916). \textbf{Lower panels (d, e, f, g, h, l, m \& n):} O, Ne, Mg, Si, S and Fe line EW maps.}\label{fig:Figure 3}}
\end{figure*}

\begin{figure*}[]
\figurenum{4}
\begin{center}
\includegraphics[angle=0,scale=0.25]{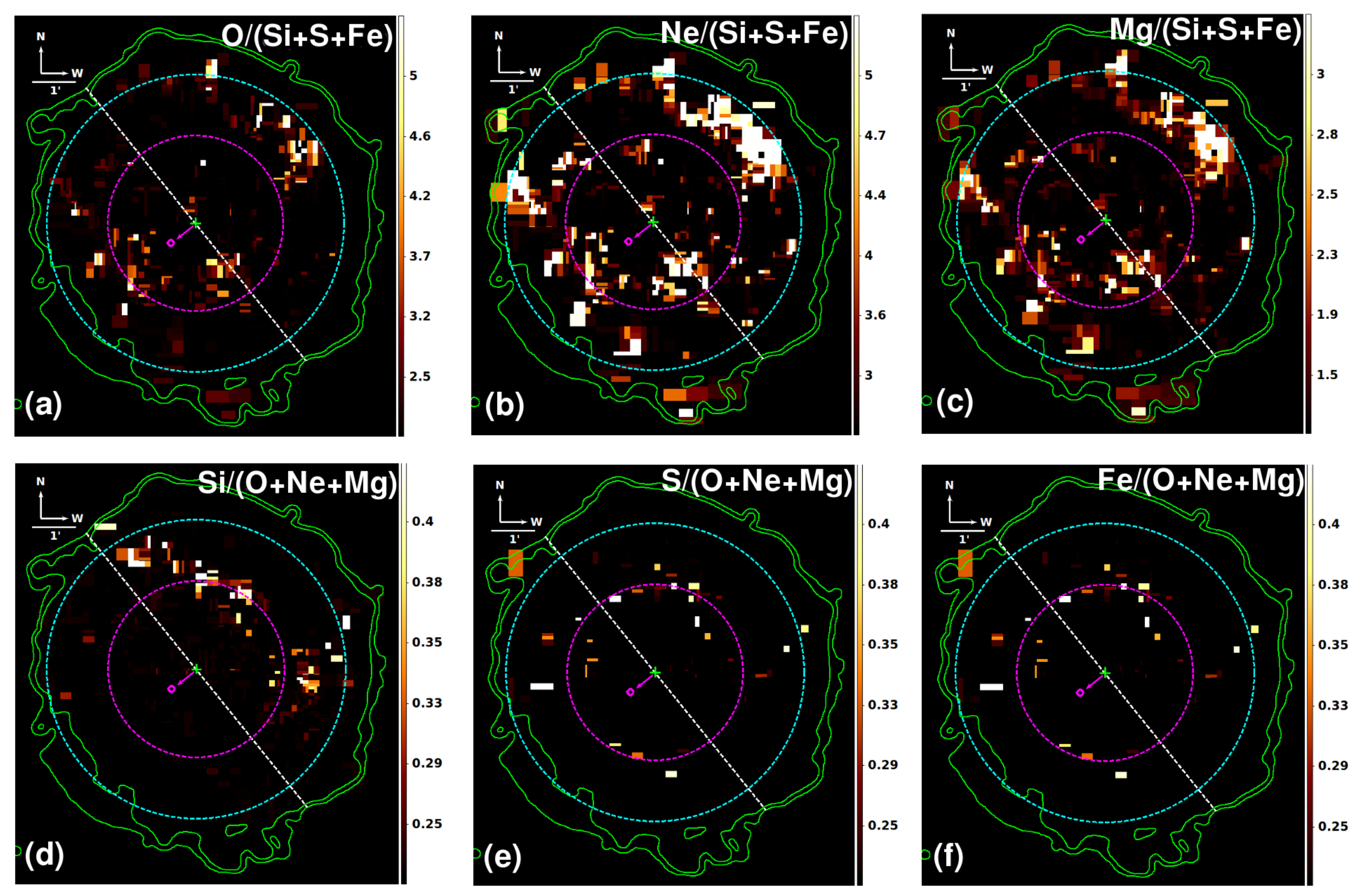}\\
\end{center}
\figcaption[]{{Abundance ratio maps for the \textit{ejecta-dominated} regions. The green cross marks the optical expansion center (\citealt{winkler2009}) and the magenta diamond marks the position of the pulsar. The magenta arrow indicates the suggested direction of the pulsar's kick. The white dashed-line is perpendicular to the direction of the pulsar's kick and divides the remnant into the northwestern and southeastern hemispheres. The dashed magenta circle represents the estimated position of the RS, and the dashed-cyan circle represents the estimated position  of the CD (\citealt{bhalerao2015}). Green contours are the outer boundary of the SNR in X-rays based on the 0.3-8 keV band ACIS-I image.}\label{fig:Figure 4}}
\end{figure*}

\begin{figure*}[]
\figurenum{5}
\begin{center}
\includegraphics[angle=0,scale=0.25]{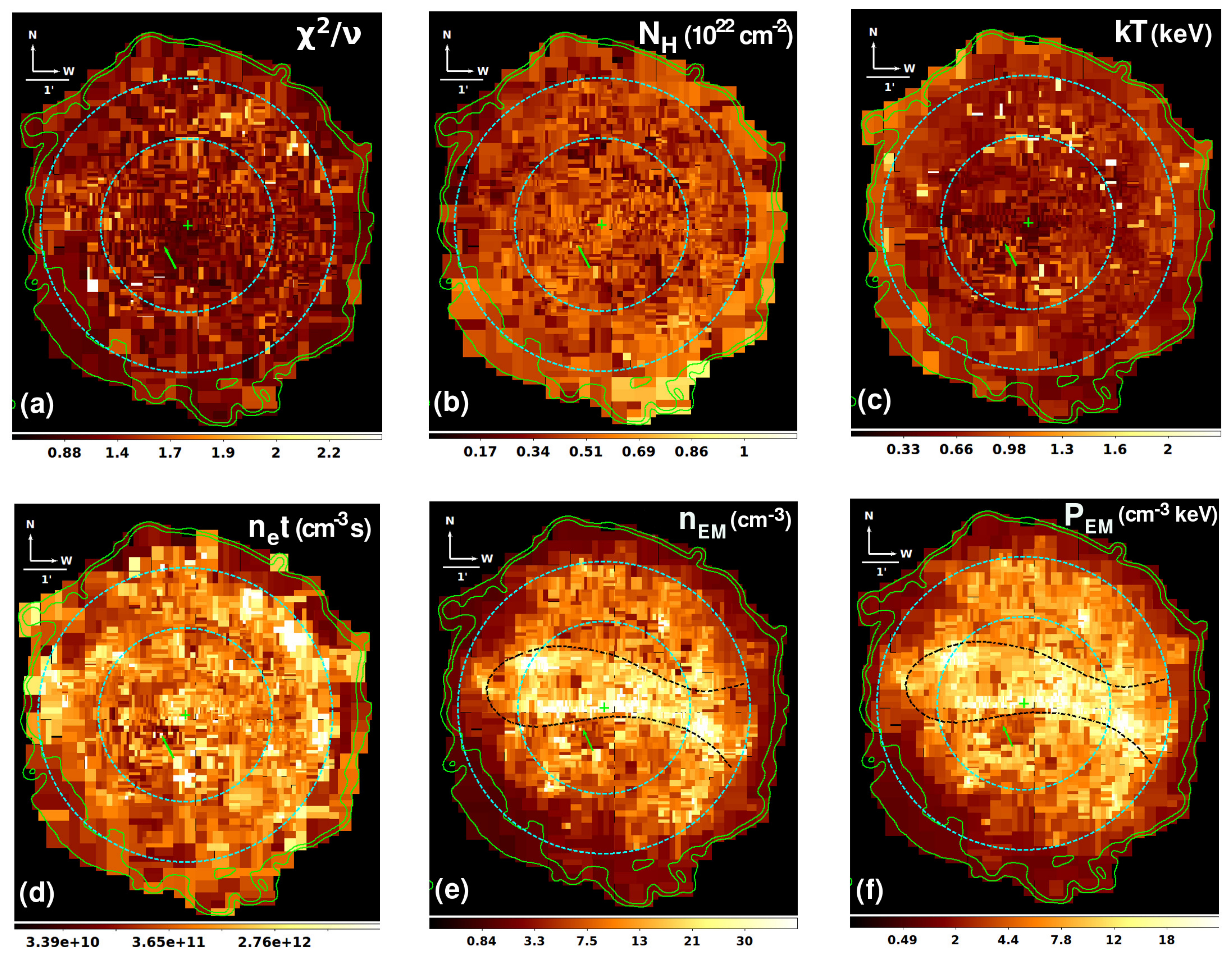}\\
\end{center}
\figcaption[]{{\textbf{(a):} Reduced chi$^2$ (\textit{$\chi^2/\nu$}), \textbf{(b):} foreground column (\textit{$N_H$}), \textbf{(c):} electron temperature (\textit{kT}), \textbf{(d):} ionization timescale (\textit{$n_e$t}), \textbf{(e)}: pseudo electron density ($n_{EM}$) and \textbf{(f):} pseudo thermal pressure ($P_{EM}$) distribution maps. For all these maps, the green contours mark the outer boundary of the SNR in the X-ray band (E = 0.3-8 keV), the green cross marks the optical expansion center, the green arrow marks the pulsar, the inner and outer cyan circles mark the estimated positions of the RS and CD respectively (\citealt{bhalerao2015}).}\label{fig:Figure 5}}
\end{figure*}

\begin{figure*}[]
\figurenum{6}
\begin{center}
\includegraphics[angle=0,scale=0.24]{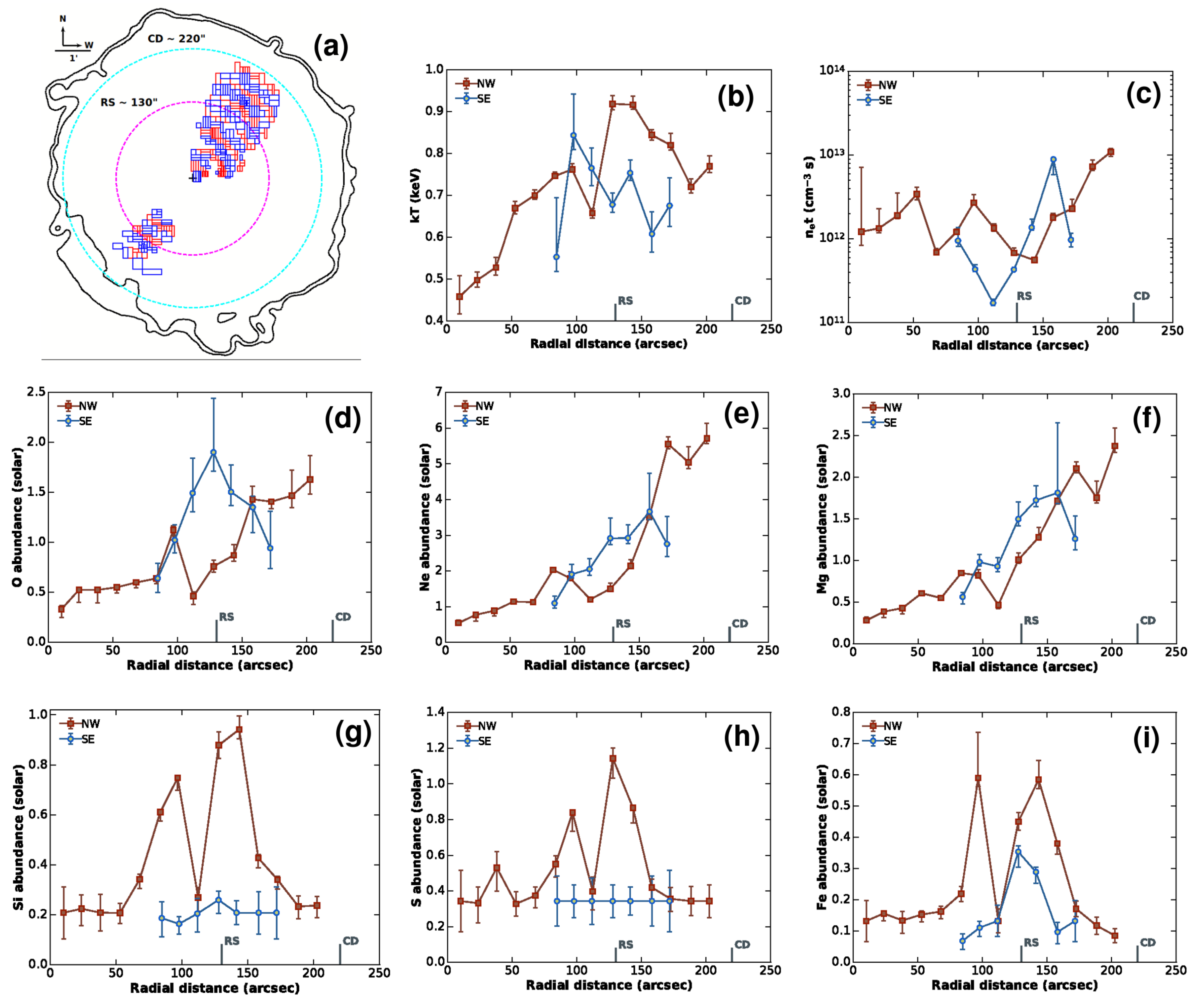}\\
\end{center}
\figcaption[]{{\textbf{(a)} Outline of G292 showing the ejecta regions used for making the radial profiles. The regions were divided into 15\arcsec~sections starting at the center and moving outwards radially. These radial slices are shown in alternating blue and red colors. Black contours are the outer boundary of the SNR in X-rays based on the 0.3-8 keV band ACIS-I image. The magenta and cyan dashed-circles represent the estimated locations of the RS and CD respectively based on the overall ejecta's kinematic structure in X-rays (\citealt{bhalerao2015}). Average values from each radial section are plotted against the distance from the expansion center (marked with a black cross). The NW regions are plotted in brown, and the SE regions in blue. The plotted parameters are: \textbf{(b)}: kT (keV),  \textbf{(c)}: n$_{e}$t (cm$^{-3}$ s), \textbf{(d)}: O, \textbf{(e)}: Ne, \textbf{(f)}: Mg, \textbf{(g)}: Si, \textbf{(h)}: S \& \textbf{(i)}: Fe abundance . The short vertical lines on the horizontal axes (labeled ``RS'' and ``CD") represent the estimated locations of the RS and CD respectively.}\label{fig:Figure 6}}
\end{figure*}

\begin{figure*}[]
\figurenum{7}
\begin{center}
\includegraphics[angle=0,scale=0.68]{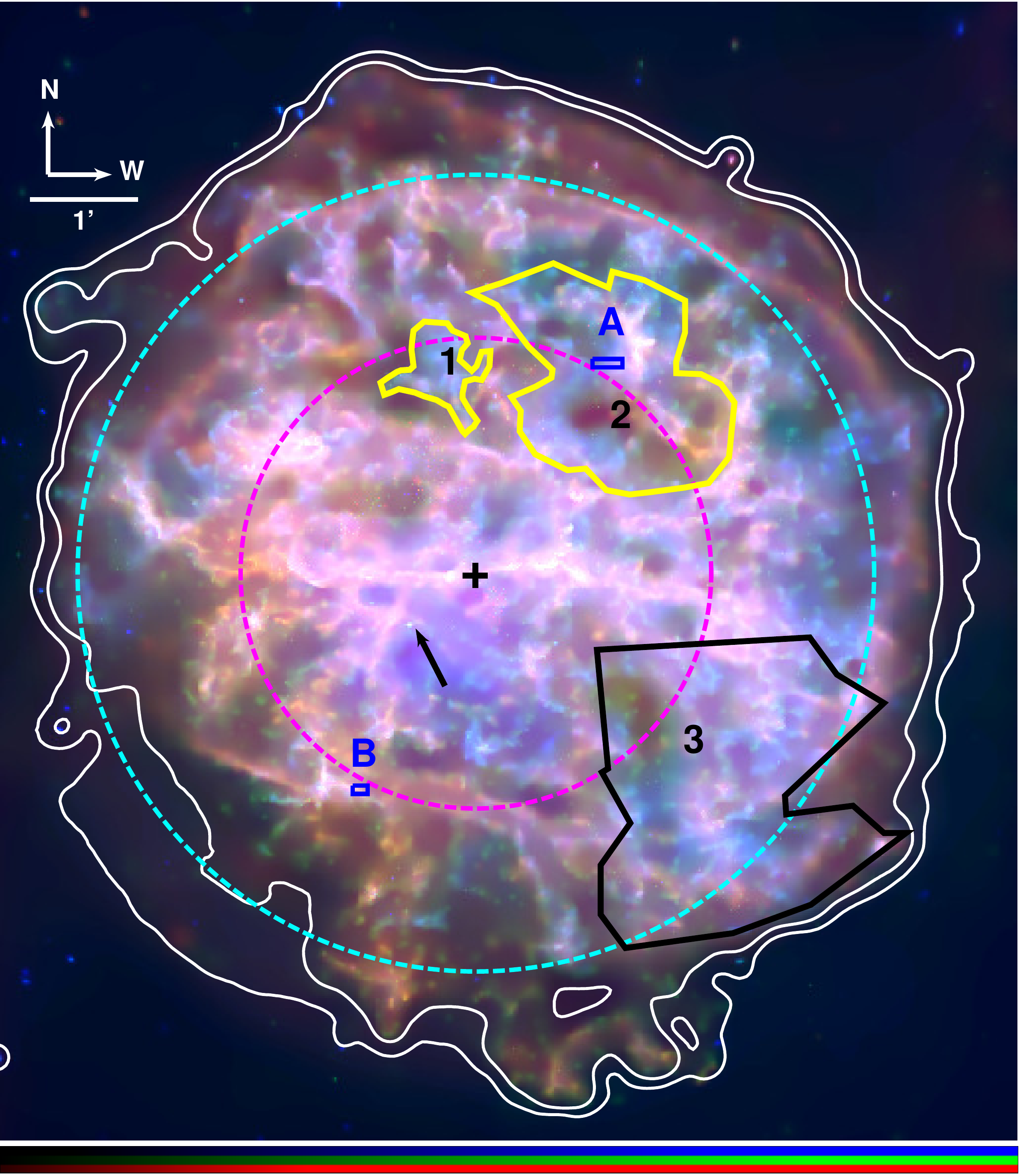}\\
\end{center}
\figcaption[]{{ACIS-I three-color image of G292 showing the extraction regions for detecting Fe emission. Red = 0.3-0.8 keV, green = 0.8-1.7 keV and blue = 1.7-8.0 keV. White contours are the outer boundary of the SNR in X-rays based on the 0.3-8 keV band ACIS-I image. The black cross and arrow mark the positions of the optical expansion center and the pulsar respectively. The dashed magenta circle represents the estimated position of the RS, and the dashed-cyan circle represents the estimated position  of the CD (\citealt{bhalerao2015}). }\label{fig:Figure 7}}
\end{figure*}

\begin{figure*}[]
\figurenum{8}
\begin{center}
\includegraphics[angle=0,scale=0.30]{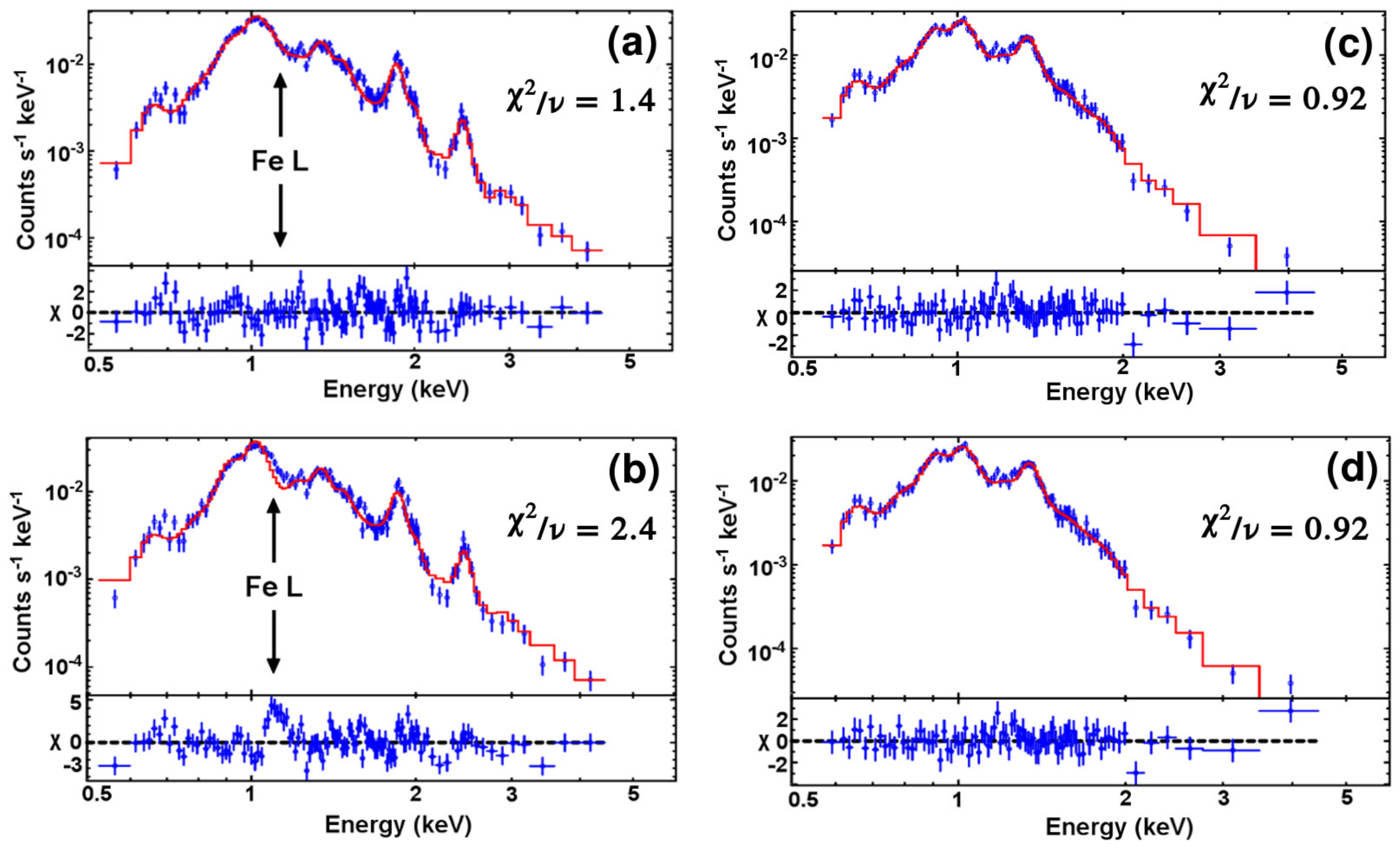}\\
\end{center}
\figcaption[]{{\textbf{(a \& b)}: Spectral model fits to the X-ray spectrum of a small Fe-rich region (region A identified in Fig. 7). In (a) the Fe abundance is varied, and the best-fit value is $9.8^{+15}_{-7.0}$$\times$ solar. In (b), for comparison, we fixed the Fe abundance at the average CSM value (0.13$\times$ solar), resulting in large residuals at $E \sim$ 1.2 keV for the Fe L line complex. The abundances of the other elements (O, Ne, Mg, Si, S, Ar and Ca) are free and the abundance of Ni is tied to that of Fe in both model fits. The red line is the fitted plane shock model, the blue markers represent data. Fixing Fe increases $\chi^{2}/\nu$ from 1.4 (a) to 2.4 (b). \textbf{(c \& d)}: Spectral model fits to the X-ray spectrum of an Fe-poor region (region B in Fig. 7). In (c) we varied the Fe abundance, and the best-fit value is $0.25^{+0.29}_{-0.14}$$\times$ solar, which is consistent with the average CSM value. In (d), for comparisons, we fixed the Fe abundance at the CSM value ( 0.13$\times$ solar), which is not statistically distinguished from the model fit shown in (c).}\label{fig:Figure 8}}
\end{figure*}

\begin{figure*}[]
\figurenum{9}
\begin{center}
\includegraphics[angle=0,scale=0.30]{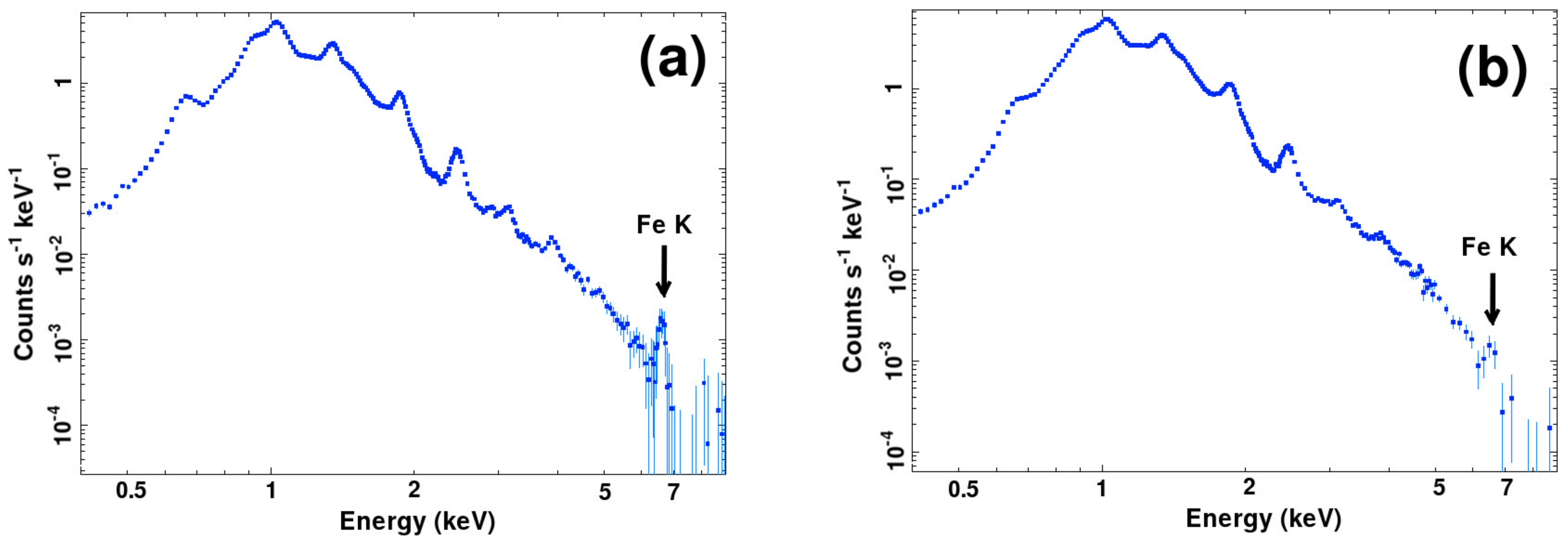}\\
\end{center}
\figcaption[]{{\textbf{(a):} X-ray spectrum extracted from the combined regions labeled 1 and 2 in Fig. 7 showing the Fe K-shell line at 6.6 keV. These two regions have $\sim$$1.2\e{6}$ counts in total. \textbf{(b)}: X-ray spectrum of an \textit{ejecta-dominated} region in the southwest (region 3 in Fig. 7) in which the Fe K-shell line detection is only marginal. This regional spectrum has $\sim$1.7\e{6} counts, which are comparable to those in Fig. 9a.}\label{fig:Figure 9}}
\end{figure*}

\begin{figure*}[]
\figurenum{10}
\begin{center}
\includegraphics[angle=0,scale=0.45]{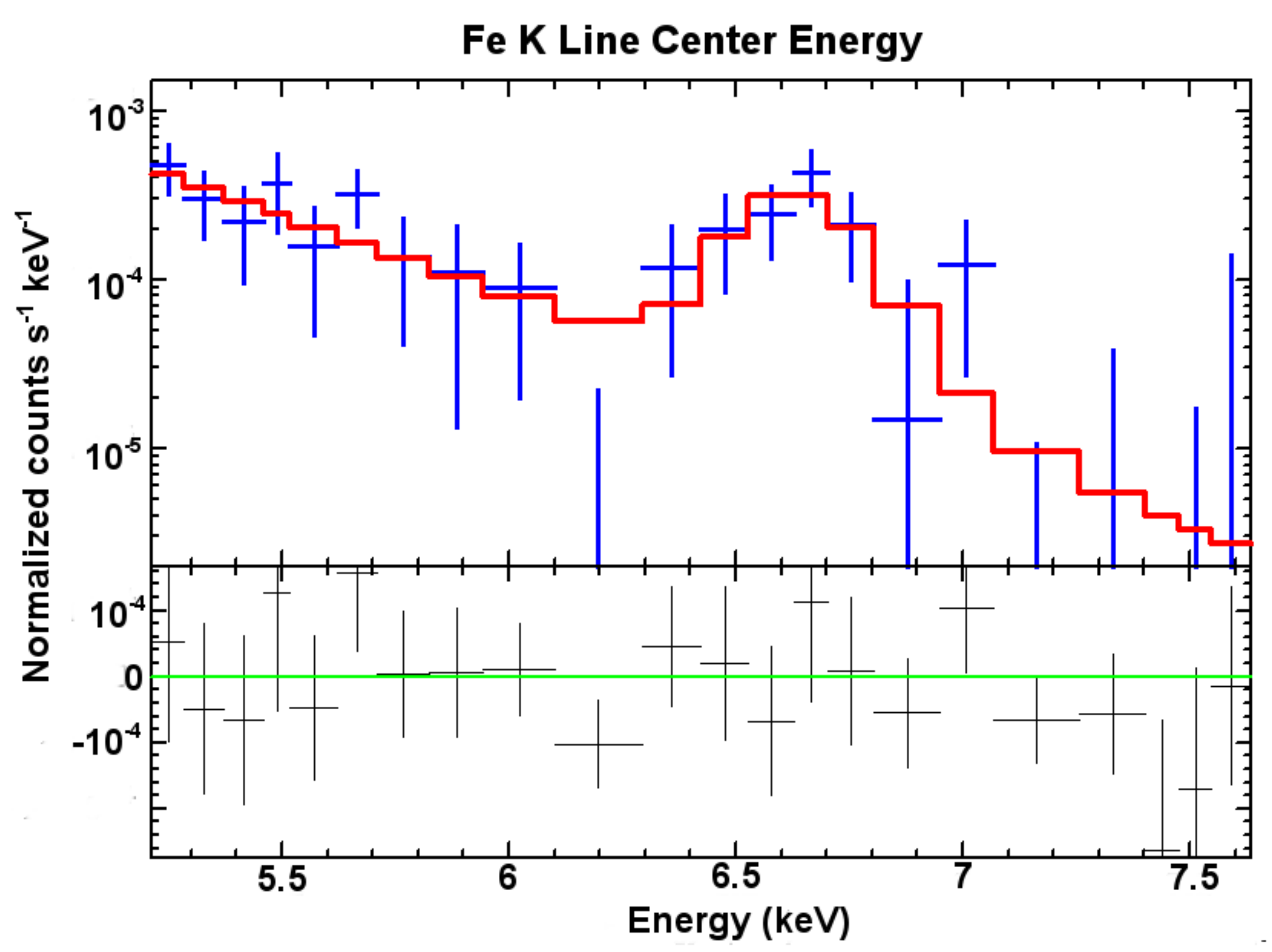}\\
\end{center}
\figcaption[]{{X-ray spectrum of 31 combined Fe-rich northern regions (seen as white boxes in Fig. 3k). The spectrum was binned at 20 counts per bin, and fitted with a Gaussian model. Data bins are in blue, and the model showing the fitted line center energy at 6.62 keV is in red. The bottom panel shows the residuals ($\chi^{2}/\nu$ $\sim$1).}\label{fig:Figure 10}}
\end{figure*}

\begin{figure*}[]
\figurenum{11}
\begin{center}
\includegraphics[angle=0,scale=0.35]{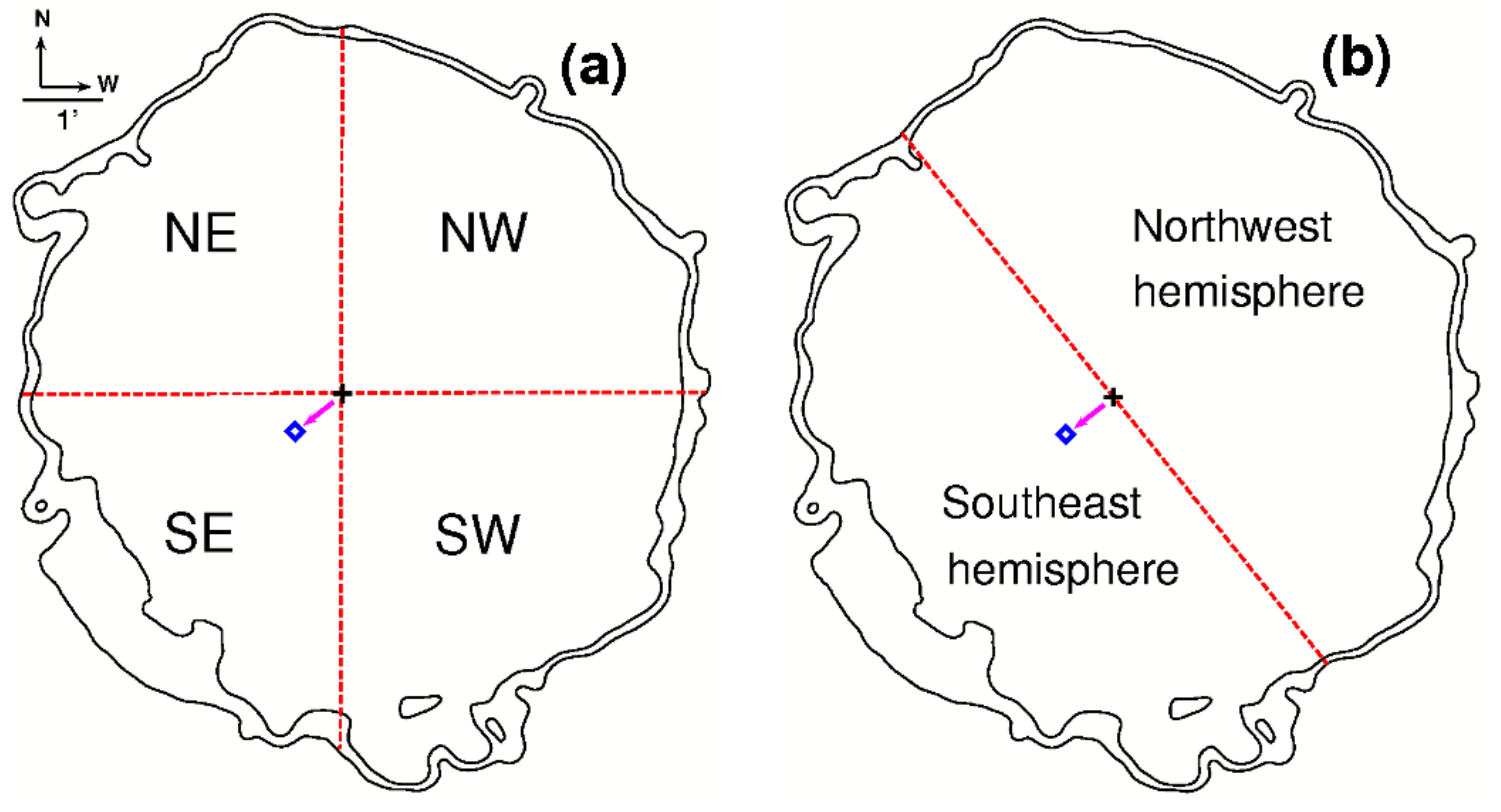}\\
\end{center}
\figcaption[]{{\textbf{(a):} Division of G292 into four quadrants: NW, NE, SE and SW \textbf{(b):} division of G292 into two hemispheres: the northwestern hemisphere, and the southeastern hemisphere to which the pulsar (PSR J1124-5916) has apparently been displaced. Black contours are the outer boundary of the SNR in X-rays based on the 0.3-8 keV band ACIS-I image. The black cross marks the position of the optical expansion center of the SNR (\citealt{winkler2009}). The location of pulsar is shown by a blue diamond and the presumed direction of the pulsar's kick is shown by a magenta arrow. The red dashed-line dividing the northwestern and southeastern hemispheres is perpendicular to the direction of the pulsar's presumed kick.} \label{fig:Figure 11}}
\end{figure*}

\begin{figure*}[]
\figurenum{12}
\begin{center}
\includegraphics[angle=0,scale=0.60]{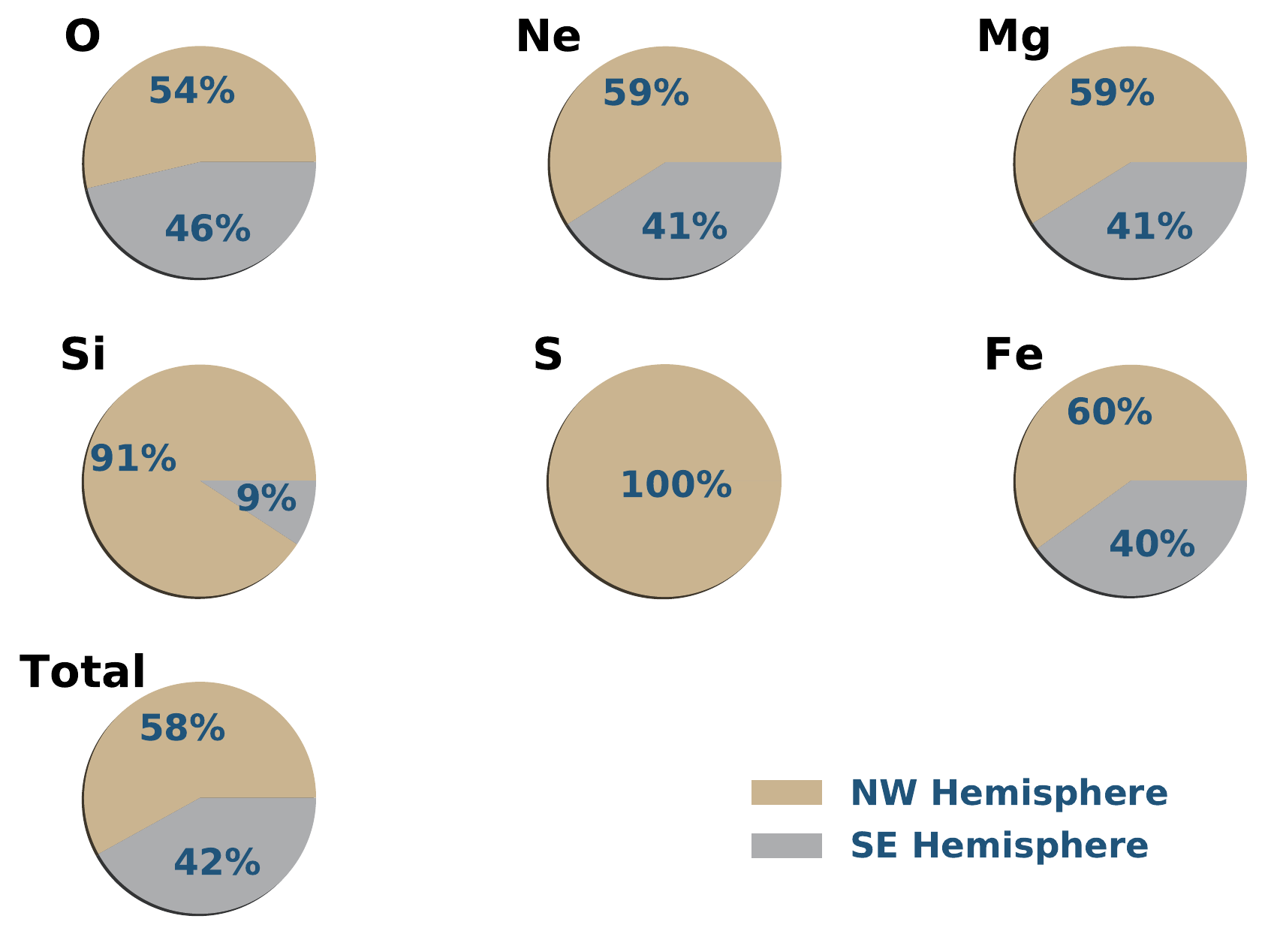}\\
\end{center}
\figcaption[]{{Pie charts showing the fractional distribution of ejecta masses for O, Ne, Mg, Si, S, Fe and the total of these six elements, between the northwestern hemisphere and the southeastern hemisphere (as shown in Fig. 11b). Ejecta regions that had the abundance of a given element greater than the CSM abundance by at least a 3$\sigma$ confidence level were considered for these plots. The northwestern hemisphere dominates the southeast especially for Si, S, and Fe, where it accounts for $\gtrsim$ 60\% of the ejecta mass. The uncertainties in the mass estimates are approximately $\pm$15\%. We note that there may be additional amounts of radiatively cooled O-, Ne-, and Mg-rich ejecta that are emitting in the optical and infrared bands but not in X-rays. These cooler ejecta have been detected mainly in a crescent-shaped structure known as the ``spur'' in the southeast (\citealt{ghavamian2005}, \citeyear{ghavamian2009}, \citeyear{ghavamian2012}; \citealt{winkler2006}; \citealt{winkler2009}). Thus, the mass-asymmetry between the NW and SE may be reduced somewhat by contributions from these cooler ejecta, however this may mainly affect the lighter O-group elements rather than the heavier Si, S, and Fe.}}\label{fig:Figure 12}
\end{figure*}

\begin{figure*}[]
\figurenum{13}
\begin{center}
\includegraphics[angle=0,scale=0.35]{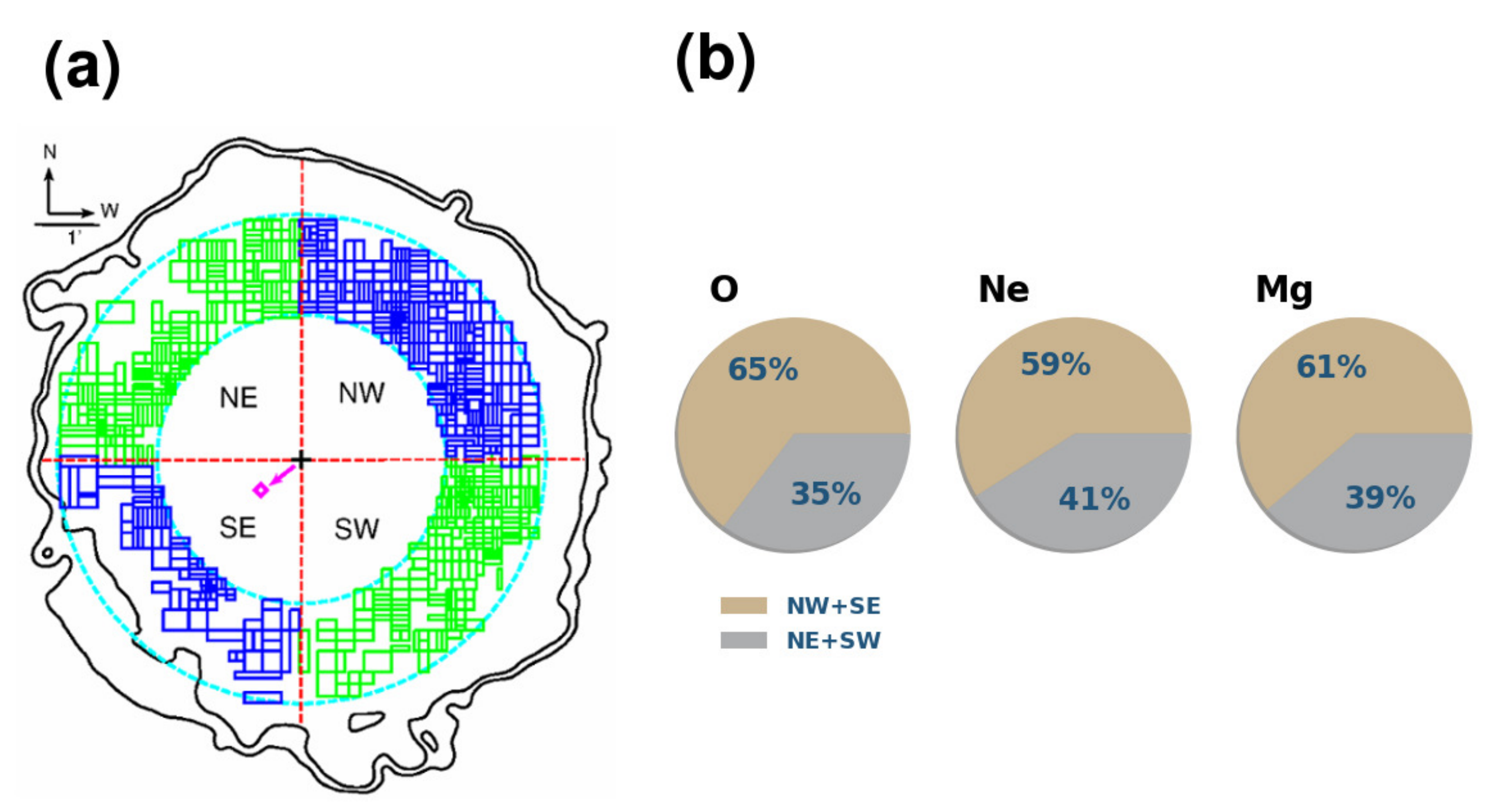}\\
\end{center}
\figcaption[]{{Pie charts showing the fractional distribution of ejecta masses for O, Ne and Mg between the combined NW and SE quadrants (blue regions in (a)), and the combined NE and SW quadrants (green regions in (a)). In order to exclude regions projected towards the central regions of the SNR, only ejecta regions between the estimated locations of the RS and CD (inner and outer dashed-cyan circles respectively) were used in this comparison. The elemental ejecta masses for these ``shell" regions are listed in Table 7. Ejecta regions that had the abundance of a given element greater than the CSM abundance by at least a 3$\sigma$ confidence level were considered in these pie charts. The uncertainties in the mass estimates are approximately $\pm$20\%. We note that contributions from cooler O-, Ne-, and Mg-rich ejecta in the southeast, that are emitting in the optical and infrared bands but not in X-rays, would further increase the mass asymmetry between the NW+SE (brown sectors in the pie charts) and NE+SW directions (gray sectors).}\label{fig:Figure 13}}
\end{figure*}

\begin{figure*}[]
\figurenum{14}
\begin{center}
\includegraphics[angle=0,scale=0.27]{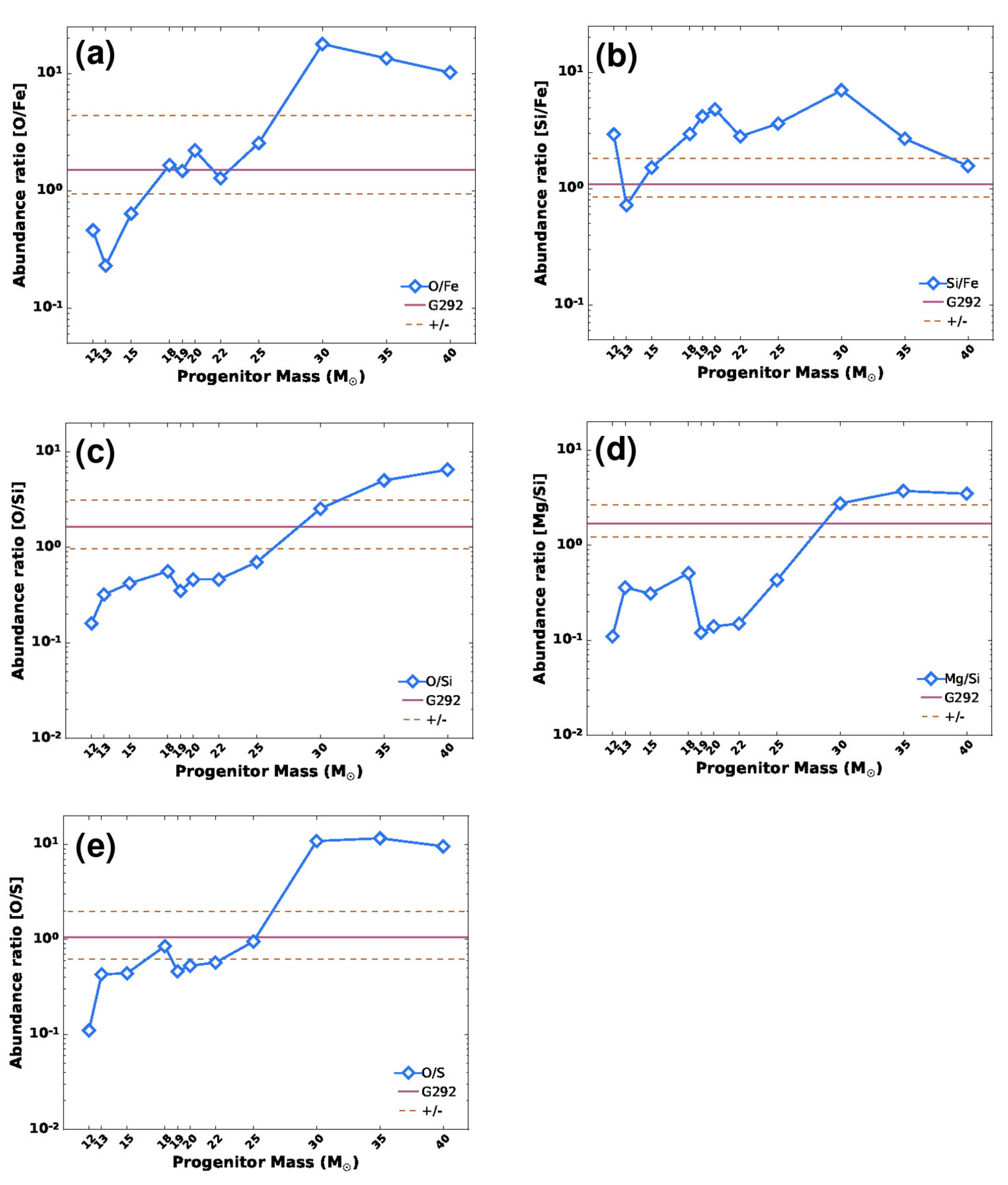}\\
\end{center}
\figcaption[]{{Average elemental abundance ratios for ejecta regions across the entire SNR (solid magenta line), compared to ratios predicted for CCSN nucleosynthesis models (blue curves, \citealt{woosley1995}). The plots represent the abundance ratios: \textbf{(a)} O/Fe, \textbf{(b)} Si/Fe and \textbf{(c)} O/Si, \textbf{(d)} Mg/Si and \textbf{(e)} O/S. The dashed lines represent the uncertainties in the measured abundance ratios for G292. Only regions with abundances $>$ 1$\times$ solar, for both the elements in a given ratio, are considered when calculating these ratios.}\label{fig:Figure 14}}
\end{figure*}

\begin{figure*}[]
\figurenum{15}
\begin{center}
\includegraphics[angle=0,scale=0.22]{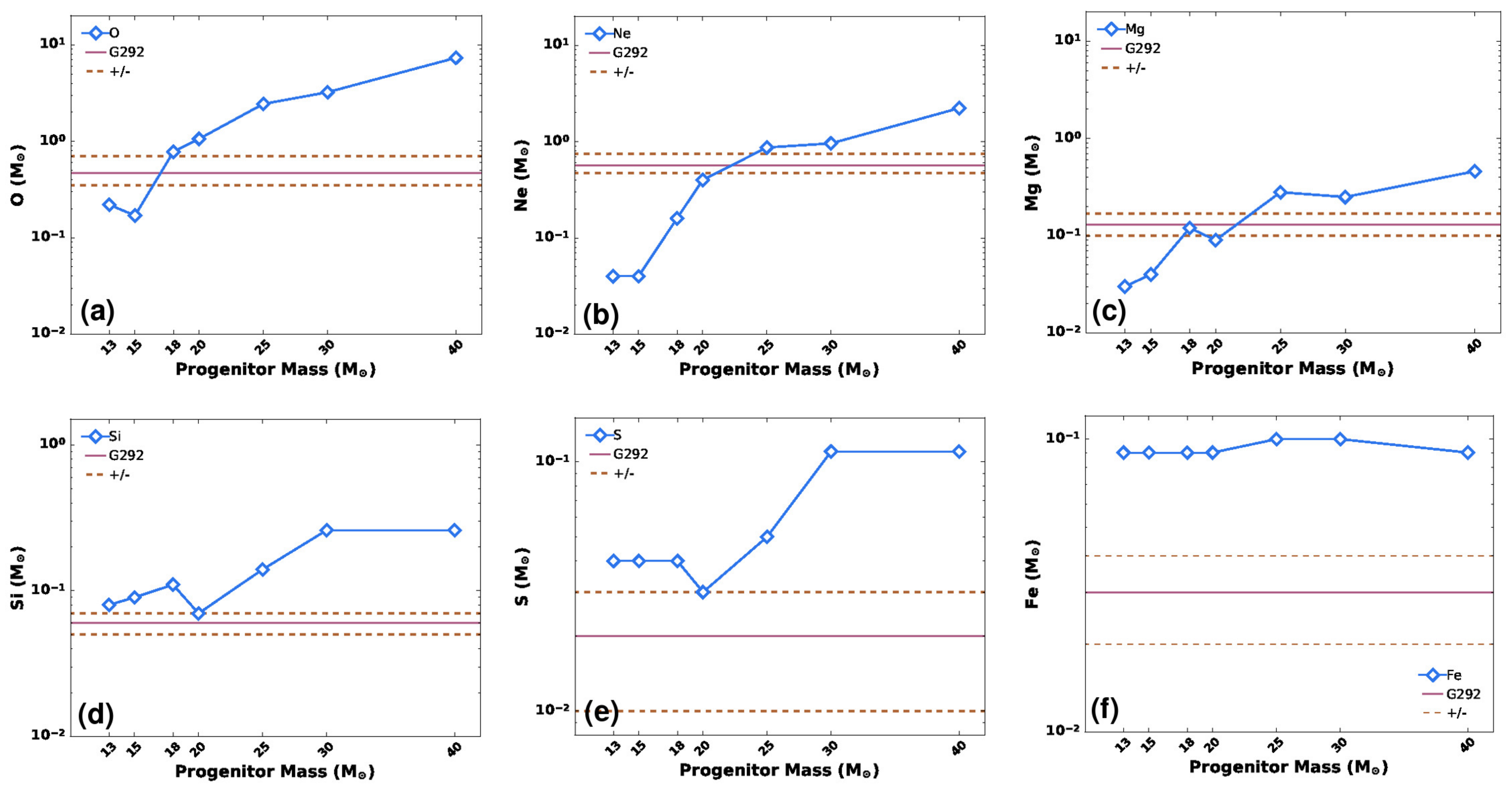}\\
\end{center}
\figcaption[]{{ Elemental ejecta masses, estimated for ejecta regions across the entire SNR (solid magenta line) compared to yields predicted for CCSN nucleosynthesis models (blue curves, $Z$ = 0.02, $E$ = 1\e{51} ergs, \citealt{nomoto2006}). The estimated ejecta masses for G292 are in terms of $f^{1/2}d^{3/2}_{6} M_{\astrosun}$, where $d_{6}$ is the distance to the SNR in units of 6 kpc. The dashed lines represent the uncertainties in the estimated ejecta masses. The plotted elemental masses are \textbf{(a)} O, \textbf{(b)} Ne, \textbf{(c)} Mg, \textbf{(d)} Si, \textbf{(e)} S and \textbf{(f)} Fe. }\label{fig:Figure 15}}
\end{figure*}

\end{document}